\documentclass{article}

\usepackage[nonatbib, final]{neurips_2020}

\usepackage[hang, flushmargin]{footmisc}

\usepackage{etoolbox} 
\makeatletter
\patchcmd\maketitle{\@makefntext}{\@@@ddt}{}{}
\patchcmd\maketitle{\rlap}{\mbox}{}{}
\makeatother

\usepackage[utf8]{inputenc} 
\usepackage[T1]{fontenc}    
\usepackage{hyperref}       
\usepackage{url}            
\usepackage{booktabs}       
\usepackage{amsfonts}       
\usepackage{nicefrac}       
\usepackage{microtype}      
\usepackage{xfrac} 

\usepackage{wrapfig}
\usepackage[percent]{overpic}
\usepackage{amsmath}        
\usepackage{enumitem}
\usepackage{graphicx}
\usepackage{tikz}
\usepackage{tikz-3dplot}
\usepackage{listings}
\usepackage{blindtext}
\usepackage{wrapfig}
\usepackage{enumitem}
\usepackage{xcolor,colortbl}
\usepackage{bm}
\usepackage{subcaption}
\usepackage{xspace}
\usepackage{placeins}
\usepackage[export]{adjustbox}

\newcommand{\lR}{\lambda}
\newcommand{\lD}{\beta}
\newcommand{\lP}{k_P}
\newcommand{\lMSE}{k_M}

\newcommand{\lRa}{\lR^{(a)}}
\newcommand{\lRb}{\lR^{(b)}}
\newcommand{\rtarget}{r_t}

\newcommand{\name}{HiFiC\xspace}
\newcommand{\nameours}{\name (Ours)\xspace}
\newcommand{\namelo}{HiFiC\textsuperscript{Lo}\xspace}
\newcommand{\namemi}{HiFiC\textsuperscript{Mi}\xspace}
\newcommand{\namehi}{HiFiC\textsuperscript{Hi}\xspace}
\newcommand{\nameourslo}{HiFiC\textsuperscript{Lo} (Ours)\xspace}
\newcommand{\nameoursmi}{HiFiC\textsuperscript{Mi} (Ours)\xspace}

\newcommand{\blmselpips}{Baseline (no GAN)\xspace}
\newcommand{\blminnen}{M\&S Hyperprior\xspace}
    
\newcommand{\ename}{\emph{\name}\xspace}
\newcommand{\enameours}{\emph{\nameours}\xspace}
\newcommand{\enamelo}{\emph{\namelo}\xspace}
\newcommand{\enamemi}{\emph{\namemi}\xspace}
\newcommand{\enamehi}{\emph{\namehi}\xspace}
\newcommand{\enameourslo}{\emph{\nameourslo}\xspace}

\newcommand{\eblmselpips}{\emph{\blmselpips}\xspace}
\newcommand{\eblminnen}{\emph{\blminnen}\xspace}
\newcommand{\million}[2]{$#1\,#200\,000$}

\definecolor{nicegray}{gray}{0.95}
\newcommand{\markgray}[1]{{\cellcolor{lightgray}{#1}}}

\newcommand*{\eg}{\textit{e.g.}}
\newcommand*{\ie}{\textit{i.e.}}
\newcommand*{\etal}{\textit{et al.}\xspace}

\newcommand{\vsqueeze}{\vspace{-.5em}}
\newcommand{\vsqueezehalf}{\vspace{-.25em}}

\hypersetup{
    colorlinks = true,
    linkcolor = magenta,
    citecolor = magenta,
}

\newcommand*\cubical[1]{
\node at (2,0,-0.6) {$C$};
\node at (4,2.1,-0.6) {$N$};
\node[rotate=90] at (-0.55,0,1.75) {$H,W$};
\node (center) at (2,2,6.5) {#1};

\foreach \i in{0,...,4}
{   
  \draw (0,0,\i) -- (4,0,\i);
  \draw (\i,0,0) -- (\i,0,4);

  \draw (4,\i, 0) -- (4, \i, 4);
  \draw (4,0, \i) -- (4, 4, \i);

  \draw (\i, 0,4) -- (\i, 4, 4);
  \draw (0, \i,4) -- (4, \i, 4);
}
}

\definecolor{normhighlight}{HTML}{B78DAD}

\newcommand{\normcube}[2]{%
\resizebox{0.175\textwidth}{!}{%
\begin{tikzpicture}[tdplot_main_coords,scale=0.4]
\draw[fill=normhighlight] #2;
\cubical{#1}
\end{tikzpicture}%
}}

\newcommand{\cropinside}[5][]{%
    \node[draw=white!40, very thick, inner sep = 0pt,anchor=north east,outer sep=1pt,#1,yshift=-1.5pt] (#4) at (#5) {%
        \includegraphics[width=0.2\linewidth]{#3}};
    \node[fill=white,anchor=south west,inner sep=1pt,outer sep=1pt, minimum height=10pt,yshift=-0pt] at (#4.south west) {#2};}

\newcommand{\fullshardlabel}[1]{\phantom{\textsuperscript L}#1\phantom{\textsuperscript L}}

\newcommand{\fullshardraw}[8][]{%
    \node[inner sep = 0pt,outer sep=0pt,anchor=south west] (image) at (0,0) {%
            \includegraphics[width=1\linewidth,#1]{#2}};
    \begin{scope}[x={(image.south east)},y={(image.north west)}]
        \def\temp{#7} \ifx\temp\empty 
        \else \node[fill=white,anchor=north west,inner sep=1pt,outer sep=0pt, minimum height=10pt] at (0,1) {\fontsize{4}{5}{#7}};
        \fi
        \draw[white, very thick, outer sep=2pt] (#5,0) -- (#6,1);
        \node[fill=white,anchor=south east,inner sep=1pt,outer sep=0pt, minimum height=10pt, xshift=-5pt] at (#5,0) {\fullshardlabel{#4}};
        \begin{scope}
            \clip (#5,0) -- (#6,1) -- (1,1) -- (1,0) -- cycle;
            \node[anchor=north east,inner sep=0pt, outer sep=0pt] at (1,1) {\includegraphics[width=1\linewidth]{#3}};
        \end{scope}
        \node[fill=white,anchor=south west,inner sep=1pt,outer sep=0pt, minimum height=10pt, xshift=5pt] at (#5,0) {\fullshardlabel{Original}};
        #8
    \end{scope}}

\makeatletter
\renewcommand{\paragraph}{%
  \@startsection{paragraph}{4}%
  {\z@}{0.5ex \@plus 0.25ex \@minus 0.2ex}{-1em}%
  {\normalfont\normalsize\bfseries}%
}
\makeatother

\renewcommand{\footnoterule}{%
  \kern -3pt
  \hrule width \textwidth height 0.5pt
  \kern 2pt
}

\title{High-Fidelity Generative Image Compression}

\author{%
  Fabian Mentzer\thanks{\hspace{1ex}{\footnotesize Work done while interning at Google Research.}\hfill Project page and demo: \href{https://hific.github.io}{\textbf{\texttt{hific.github.io}}} } \\
  ETH Z\"urich
  \And
  George Toderici \\
  Google  Research 
  \And
  Michael Tschannen\thanks{\hspace{1ex}{\small Work done while at Google Research.\hfill Correspondence: \texttt{eirikur@google.com}}} \\
  \And
  Eirikur Agustsson \\
  Google Research 
}

\begin{document}

\maketitle

\begin{abstract}
We extensively study how to combine Generative Adversarial Networks and learned compression to obtain a state-of-the-art generative lossy compression system. In particular, we investigate normalization layers, generator and discriminator architectures, training strategies, as well as perceptual losses. In contrast to previous work, i) we obtain visually pleasing reconstructions that are perceptually similar to the input, ii) we operate in a broad range of bitrates, and iii) our approach can be applied to high-resolution images. We bridge the gap between rate-distortion-perception theory and practice by evaluating our approach both quantitatively with various perceptual metrics, and with a user study. The study shows that our method is preferred to previous approaches even if they use more than $2{\times}$ the bitrate.
\end{abstract}

\section{Introduction}

The ever increasing availability of cameras produces an endless stream of images. To store them efficiently, lossy image compression algorithms are used in many applications. Instead of storing the raw RGB data, a lossy version of the image is stored, with---hopefully---minimal visual changes to the original. Various algorithms have been proposed over the years~\cite{jpeg1992wallace,jpeg2000taubman,webpurl},
including using state-of-the-art video compression algorithms for single image compression (BPG~\cite{bpgurl}). 
At the same time, deep learning-based lossy compression has seen great interest~\cite{toderici2015variable,balle2016end,mentzer2018conditional}, where a neural network is directly optimized for the rate-distortion trade-off, which led to new state-of-the-art methods.

However, all of these approaches degrade images significantly as the compression factor increases.
While classical algorithms start to exhibit algorithm-specific artifacts such as blocking or banding, learning-based approaches reveal issues with the distortion metric that was used to train the networks.
Despite the development of a large variety of ``perceptual metrics'' (\eg, (Multi-Scale) Structural Similarity Index (\mbox{(MS-)SSIM)}~\cite{SSIM,SSIM-MS}, Learned Perceptual Image Patch Similarity (LPIPS)~\cite{zhang2018unreasonable}), the weakness of each metric is exploited by the learning algorithm, \eg, checkerboard artifacts may appear when targeting neural network derived metrics, relying on MS-SSIM can cause poor text reconstructions, and MSE yields blurry reconstructions.

In~\cite{agustsson2019extreme}, Agustsson~\etal demonstrated the potential of using GANs to prevent compression artifacts with a compression model that produces perceptually convincing reconstructions for extremely low bitrates (${<}0.08$ bpp).
However, their reconstructions tend to only preserve high-level semantics, deviating significantly from the input.

Recent work by Blau and Michaeli~\cite{blau2019rethinking} characterized this phenomenon by showing the existence of a triple ``rate-distortion-perception'' trade-off, formalizing ``distortion'' as a similarity metric comparing pairs of images, and ``perceptual quality'' as the distance between the image distribution $p_X$ and the distribution of the reconstructions $p_{\hat X}$ produced by the decoder, measured as a distribution divergence. They show that at a fixed rate, better perceptual quality always implies worse distortion. Conversely, only minimizing distortion will yield poor perceptual quality.  To overcome this issue, distortion can be traded for better perceptual quality by minimizing the mismatch between the distributions of the input and the reconstruction, \eg, with Generative Adversarial Networks (GANs)~\cite{goodfellow2014generative}. While \cite{blau2019rethinking} presents comprehensive theoretical results, the rate-distortion-perception trade-off is only explored empirically on toy datasets. 

In this paper, we address these issues with the following contributions:
\begin{enumerate}[leftmargin=*,topsep=0pt]
    \item We propose a generative compression method to achieve high quality reconstructions that are very close to the input, for high-resolution images (we test up to $2000{\times}2000$ px). 
    In a user study, we show that our approach is visually preferred to previous approaches \emph{even when these approaches use more than $2{\times}$ higher bitrates}, see Fig.~\ref{fig:userstudy}. 
    \item We quantitatively evaluate our approach with FID~\cite{heusel2017gans}, KID~\cite{binkowski2018demystifying}, NIQE~\cite{mittal2012making}, LPIPS~\cite{zhang2018unreasonable}, and the classical distortion metrics PSNR, MS-SSIM, and show how our results are consistent with the rate-distortion-perception theory. We also show that no metric would have predicted the full ranking of the user study, but that FID and KID are useful in guiding exploration.
    Considering this ensemble of diverse perceptual metrics including no-reference metrics, pair-wise similarities, and distributional similarities, as well as deep feature-based metrics derived from different network architectures, ensures a robust perceptual evaluation.
    \item We extensively study the proposed architecture and its components, including normalization layers, generator and discriminator architectures,  training strategies, as well as the loss, in terms of perceptual metrics and stability.
\end{enumerate}

\begin{figure}
\tiny
  \begin{tikzpicture}
    \fullshardraw[trim={0 9.5cm 0 0},clip]{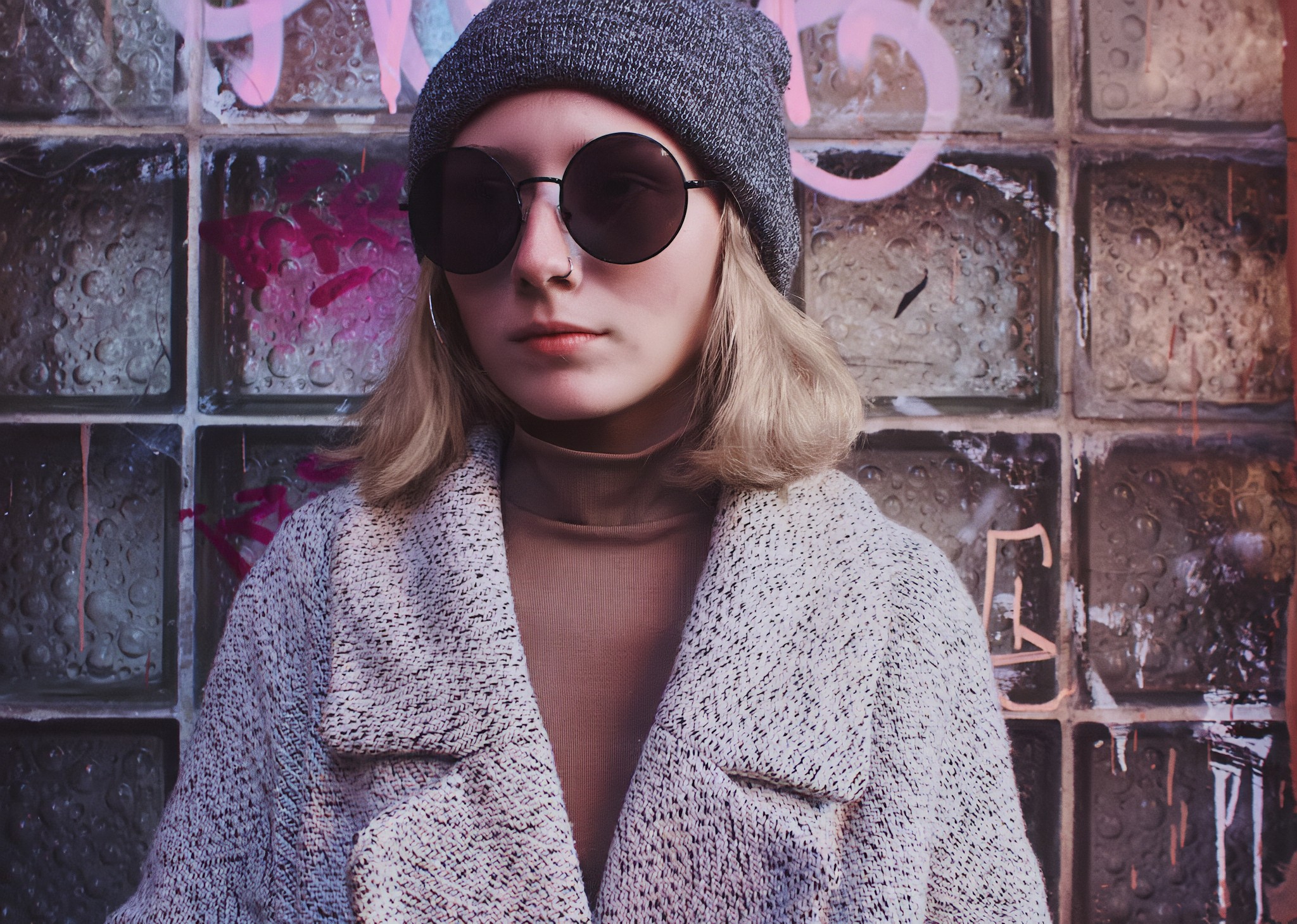}{%
                  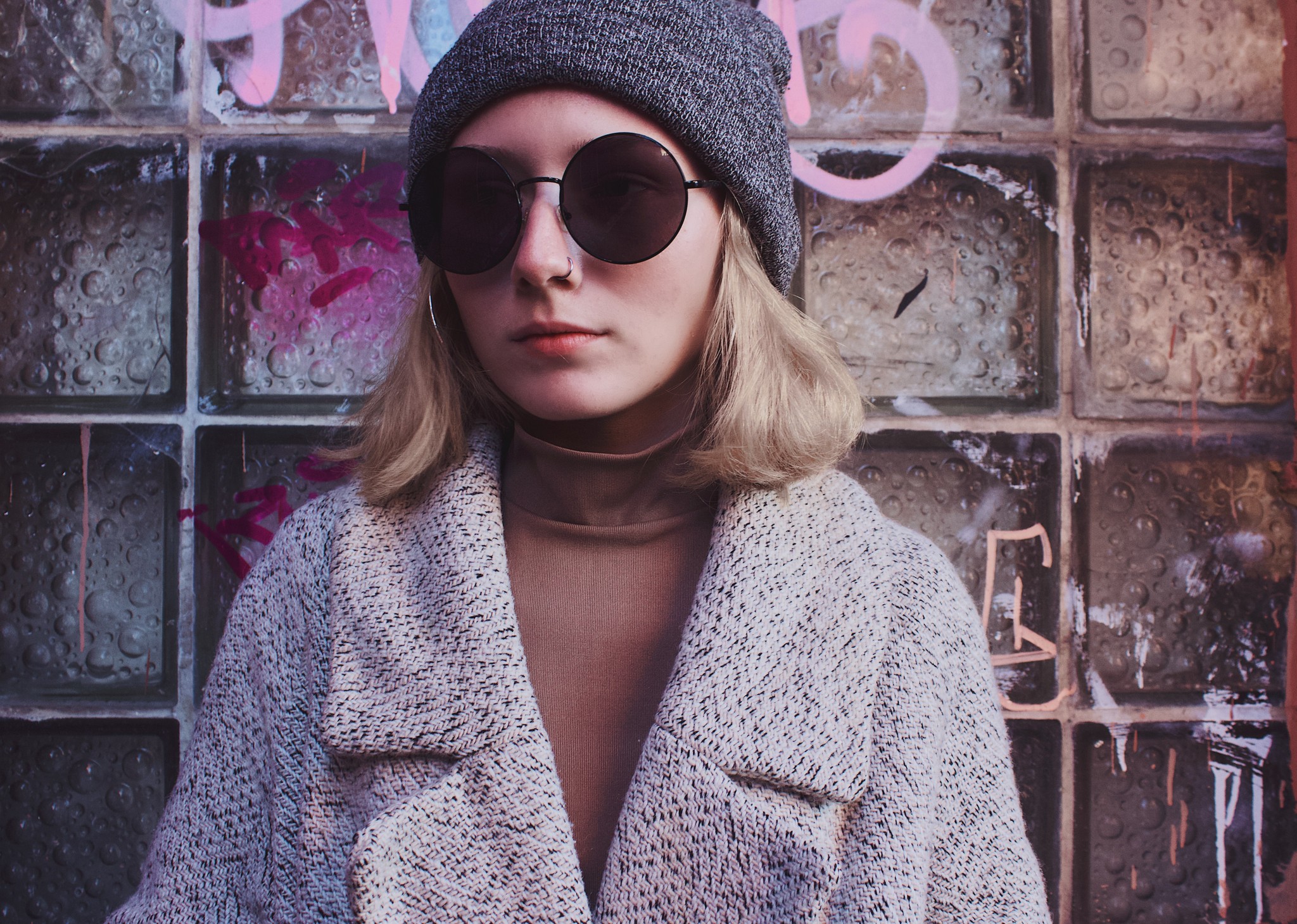}{%
                  \textbf{\nameourslo}: 0.198bpp}{0.45}{0.45}{}{
        \cropinside[xshift=-1.5pt]{Original}{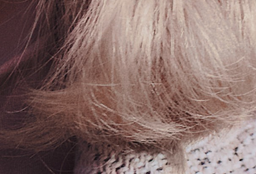}{cropra}{1,1}
        \cropinside{\textbf{\namelo}: 0.198bpp}{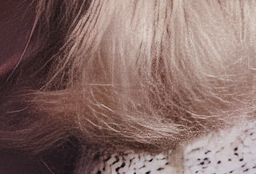}{croprb}{cropra.south east}
        \cropinside{BPG: 0.224bpp}{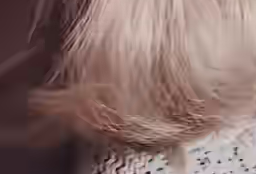}{croprc}{croprb.south east}
        \cropinside{BPG: 0.446bpp}{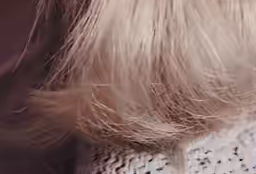}{croprd}{croprc.south east}
        \cropinside[anchor=north west,xshift=1.5pt]{Original}{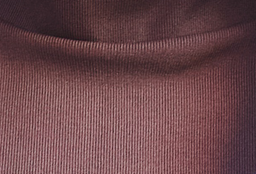}{cropla}{0,1}
        \cropinside{\textbf{\namelo}: 0.198bpp}{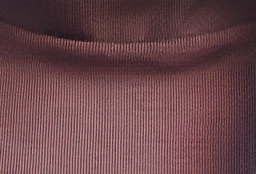}{croplb}{cropla.south east}
        \cropinside{BPG: 0.224bpp}{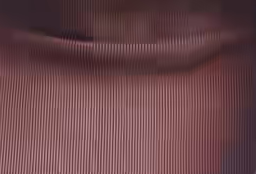}{croplc}{croplb.south east}
        \cropinside{BPG: 0.446bpp}{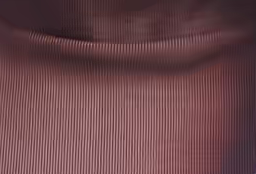}{cropld}{croplc.south east}}
  \end{tikzpicture}
\vspace{-4ex}
\caption{\label{fig:frontpic}Comparing our method, \ename, to the original, as well as BPG at a similar bitrate and at $2{\times}$ the bitrate. We can see that our GAN model produces a high-fidelity reconstruction that is very close to the input, while BPG exhibits blocking artifacts, that are still present at double the bitrate. In the background, we show a split to the original to further indicate how close our reconstruction is. 
We show many more visual comparisons in Appendix~\ref{sec:supp:morevisuals}, including more methods, more bitrates, and various datasets. \hfill \emph{Best viewed on screen.}\vspace{-2ex}}
\end{figure}

\section{Related work}

The most frequently used lossy compression algorithm is JPEG~\cite{jpeg1992wallace}. Various hand-crafted algorithms have been proposed to replace JPEG, including WebP~\cite{webpurl} and JPEG2000~\cite{jpeg2000taubman}. 
Relying on the video codec HEVC~\cite{sullivan2012overview}, BPG~\cite{bpgurl} achieves very high PSNR across varying bitrates. 
Neural compression approaches directly optimize Shannon's rate-distortion trade-off~\cite{cover2012elements}. Initial works relied on RNNs~\cite{toderici2015variable,toderici2016full}, while subsequent works were based on auto-encoders~\cite{balle2016end,theis2017lossy,agustsson2017soft}. To decrease the required bitrate, various approaches have been used to more accurately model the probability density of the auto-encoder latents for more efficient arithmetic coding, using hierarchical priors, auto-regression with various context shapes, or a combination thereof~\cite{balle2018variational,mentzer2018conditional,li2017learning,rippel17a,minnen2018joint,lee2018context, minnen2020channel}. State-of-the-art models now outperform BPG in PSNR, \eg, the work by Minnen \etal~\cite{minnen2018joint}.

Since their introduction by Goodfellow~\etal \cite{goodfellow2014generative}, GANs have led to rapid progress in unconditional and conditional image generation. State-of-the-art GANs can now generate photo-realistic images at high resolution \cite{brock2018large, karras2019style, park2019semantic}. Important drivers for this progress were increased scale of training data and model size \cite{brock2018large}, innovation in network architectures \cite{karras2019style}, and new normalization techniques to stabilize training \cite{miyato2018spectral}. Beyond (un)conditional generation, adversarial losses led to advances in different image enhancement tasks, such as compression artifact removal \cite{galteri2017deep}, image de-noising \cite{chen2018image}, and image super-resolution \cite{ledig2017photo}. Furthermore, adversarial losses were previously employed to improve the visual quality of neural compression systems \cite{rippel17a, santurkar2017generative, tschannen2018deep, agustsson2019extreme, blau2019rethinking}. \cite{rippel17a} uses an adversarial loss as a component in their full-resolution compression system, but they do not systematically ablate and asses the benefits of this loss on the quality of their reconstructions. While \cite{santurkar2017generative} provides a proof-of-concept implementation of a low-resolution compression system with a GAN discriminator as decoder, \cite{tschannen2018deep, blau2019rethinking} focus on incorporating a GAN loss into the rate-distortion objective in a conceptually sound fashion. Specifically, \cite{tschannen2018deep} proposes to augment the rate-distortion objective with a distribution constraint to ensure that the distribution of the reconstructions match the input distribution at all rates, and \cite{blau2019rethinking} introduce and study a triple trade-off between rate, distortion, and distribution matching. Finally, \cite{agustsson2019extreme} demonstrates that using GAN-based compression systems at very low bitrates can lead to bitrate savings of $2\times$ over state-of-the-art engineered and learned compression algorithms.

\section{Method}\vspace{-.5em}
\subsection{Background}

\paragraph{Conditional GANs}
Conditional GANs~\cite{goodfellow2014generative,mirza2014conditional} are a way to learn a generative model of a conditional distribution $p_{X|S}$, where each datapoint $x$ is associated with additional information $s$ (\eg, class labels or a semantic map)
and $x, s$ are related through an unknown joint distribution $p_{X,S}$. 
Similar to standard GANs, in this setting we train two rivaling networks: 
a generator $G$ that is conditioned on $s$ to map samples $y$ from a fixed known distribution $p_Y$ to $p_{X|S}$, and a discriminator $D$ that maps an input $(x, s)$ to the probability that it is a sample from $p_{X|S}$ rather than from the output of $G$. The goal is for $G$ to ``fool'' $D$ into believing its samples are real, \ie, from $p_{X|S}$. 
Fixing $s$, we can optimize the ``non-saturating'' loss~\cite{goodfellow2014generative}:
\begin{align}
    \mathcal{L}_G &= \mathbb{E}_{y \sim p_Y} [ 
            -\log(D(G(y, s), s) ], \nonumber \\
    \mathcal{L}_D &= \mathbb{E}_{y \sim p_Y} [ 
            -\log(1 - D(G(y, s), s) ]
            + \mathbb{E}_{x \sim p_{X|s}} [ -\log(D(x, s)) ]. 
            \label{eq:nonsat}
\end{align}

\paragraph{Neural Image Compression}
Learned lossy compression is based on Shannon's rate-distortion theory~\cite{cover2012elements}. 
The problem is usually modeled with an auto-encoder consisting of an encoder $E$ and a decoder $G$. To encode an image $x$, we obtain a quantized latent $y = E(x)$. To decode, we use $G$ to obtain the lossy reconstruction $x' = G(y)$. The compression incurs a distortion $d(x, x')$, \eg, $d = \text{MSE}$ (mean squared error).
Storing $y$ is achieved by introducing a probability model $P$ of $y$. Using $P$ and an entropy coding algorithm (\eg, arithmetic coding~\cite{marpe2003context}), we can store $y$ losslessly using bitrate $r(y) = - \log(P(y))$ (in practice there is a negligible bit-overhead incurred by the entropy coder).
If we parameterize $E, G$, and $P$ as CNNs, we can train them jointly by minimizing the \emph{rate-distortion trade-off},
where $\lR$ controls the trade-off:
\begin{equation}
    \mathcal{L}_{EG} = \mathbb{E}_{x \sim p_X} [
    \lR \; r(y) + d(x, x')]. \label{eq:rd}
\end{equation}

\vspace{-.5em}
\subsection{Formulation and Optimization} \label{sec:formulation}
\vspace{-.5em}

We augment the neural image compression formulation with a conditional GAN, \ie, we merge  Eqs.~\ref{eq:nonsat},~\ref{eq:rd} and learn networks $E$, $G$, $P$, $D$.
We use $y = E(x)$, and $s = y$.
Additionally, we use the ``perceptual distortion'' $d_P=\text{LPIPS}$, inspired by~\cite{wang2018highres}, who showed that using a VGG~\cite{simonyan2014very}-based loss helps training.
In the formalism of~\cite{blau2019rethinking}, $d_P$ is a distortion (as it is applied point-wise), thus we group it together with MSE to form our distortion $d = \lMSE \text{MSE} + \lP d_P$, where $\lMSE, \lP$ are hyper-parameters. Using hyper-parameters $\lR, \lD$ to control the trade-off between the terms, we obtain:
\begin{align}
    \mathcal{L}_{EGP} &= \mathbb{E}_{x \sim p_X} [
                    \lR r(y) + 
                    d(x, x')
                    -\lD \log(D(x', y)) ], \label{eq:lossG} \\
    \mathcal{L}_D &= \mathbb{E}_{x \sim p_X} [
                        - \log(1 - D(x', y)) ]
                    + \mathbb{E}_{x \sim p_X} [
                        - \log(D(x, y)) ]. \label{eq:lossD}
\end{align}
\paragraph{Constrained Rate} When training a neural compression model w.r.t.\ the loss in Eq.~\ref{eq:rd}, there is only a single term, $d(x,x')$, that is at odds with the rate term $r(y)$. 
The final (average) bitrate of the model can thus be controlled by varying only $\lR$.
In our setting however, MSE, $d_P$, and $-\log(D(x'))$ are at odds with the rate. For a fixed $\lR$, different $\lMSE, \lP, \lD$ would thus result in models with different bitrates, making comparison hard. To alleviate this, we introduce a ``rate target'' hyper-parameter $\rtarget$, and replace $\lR$ in Eq.~\ref{eq:lossG} with an \emph{adaptive} term $\lR'$ that uses the two hyper parameters $\lRa, \lRb$, where $\lR' = \lRa$ if $r(y) > \rtarget$, and $\lR' = \lRb$ otherwise.
Setting $\lRa \gg \lRb$ allows us to learn a model with an average bitrate close to $r_t$.

\subsection{Architecture} \label{sec:arch}

\begin{figure}
    \centering
    \includegraphics[width=\textwidth]{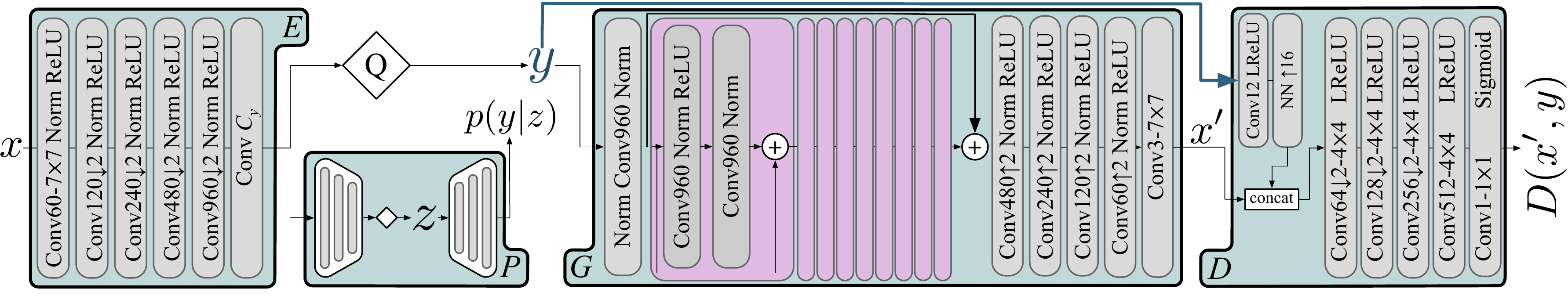}
    \vspace{-4ex}
    \caption{Our architecture. \emph{ConvC} is a convolution with $C$ channels, with $3{\times}3$ filters, except when denoted otherwise. $\downarrow2$, $\uparrow2$ indicate strided down or up convolutions. \emph{Norm} is ChannelNorm (see text), \emph{LReLU} the leaky ReLU~\cite{xu2015empirical} with $\alpha{=}0.2$, 
    \emph{NN${\uparrow}16$} nearest neighbor upsampling,
    $Q$ quantization.}
    \vspace{-3.5ex}
    \label{fig:arch}
\end{figure}

We show our architecture in Fig.~\ref{fig:arch}, including the encoder $E$, generator $G$, discriminator $D$ and probability model $P$.
For $P$, we use the hyper-prior model proposed in~\cite{balle2018variational}, where we extract side information $z$ to model the distribution of $y$ and simulate  quantization with uniform noise $\mathcal{U}(-1/2, 1/2)$ in the hyper-encoder and when estimating $p(y|z)$.
However, when feeding $y$ to $G$ we use rounding instead of noise (\ie, the straight-through estimator as in \cite{theis2017lossy}), which ensures that $G$ sees the same quantization noise during training and inference.
$E, G$ and $D$ are based on~\cite{wang2018highres,agustsson2019extreme}, with some key differences in the discriminator and in the normalization layers, described next.

Both~\cite{wang2018highres,agustsson2019extreme} use a multi-scale patch-discriminator $D$, while we only use a single scale, and we replace InstaceNorm~\cite{ulyanov2016instance} with SpectralNorm~\cite{miyato2018spectral}. Importantly, and in contrast to~\cite{agustsson2019extreme}, we condition $D$ on $y$ by concatenating an upscaled version to the image, as shown in Fig.~\ref{fig:arch}. This is motivated from using a conditional GAN formulation, where $D$ has access to the the conditioning information (which for us is $y$, see Section~\ref{sec:formulation}).

In~\cite{wang2018highres} InstanceNorm is also used in $E, G$. In our experiments, we found that this introduces significant darkening artifacts when employing the model on images with a different resolution than the training crop size (see Section~\ref{sec:results:quant}).
We assume this is due to the spatial averaging of InstanceNorm, and note that~\cite{park2019semantic} also saw issues with InstanceNorm. To alleviate these issues, we introduce \textbf{ChannelNorm}, which normalizes over channels. 
The input is a batch of $C{\times}H{\times}W$-dimensional feature maps $f_{chw}$, which are normalized to 
\begin{align}
    f'_{chw} = \frac{f_{chw}-\mu_{hw}}{\sigma_{hw}}
                \alpha_c + \beta_c, \quad \text{where} \quad
        &\mu_{hw} = \sfrac{1}{C} \textstyle \sum_{c=1}^C f_{chw} \\ \nonumber
        &\sigma_{hw}^2 = \sfrac{1}{C} \textstyle \sum_{c=1}^C (f_{chw} - \mu_{hw})^2,
\end{align}
using learned per-channel offsets $\alpha_c, \beta_c$.
We note that ChannelNorm has similarities to other normalizations. 
Most closely related, Positional Normalization~\cite{li2019positionalnorm} also normalizes over channels, but does not use learned $\alpha_c, \beta_c$. Instead, they propose to use $\mu, \sigma$ from earlier layers as offsets for later layers, motivated by the idea that these $\mu, \sigma$ may contain useful information.
Further, Kerras~\etal~\cite{karras2017progressive} normalize each feature vector ``pixelwise'' to unit length, without centering ($\mu=0$).
We show a visualization and more details in Appendix~\ref{sec:channelnormdetails}.

\subsection{User Study} \label{sec:userstudy}

To visually validate our results, we set up a user study as two-alternative forced choice (2AFC) on $N_I{=}20$ random images of the CLIC2020~\cite{clic2020} dataset (datasets are described below), inpsired by the methodology used in~\cite{clic2020}.
Participants see crops of these images, which are determined as follows: when an image is first shown, a random $768{\times}768$ crop is selected. The participants are asked to request a new random crop until they find ``an interesting crop'' (\ie, a region that is not completely flat or featureless), and then they use this crop for all comparisons on that image. We chose the crop size such that it is as large as possible while still allwoing two crops to fit side-by-side on all screens, downsampling would cause unwanted biases. The crops selected by users are linked in Appendix~\ref{sec:supp:userstudy}.  

At any time, the participants are comparing a pair of methods A and B. On screen, they always see two crops, the left is either method A or B, the right is always the original (Fig.~\ref{fig:userstudy_gui} shows a screenshot). 
Participants are asked to toggle between methods to ``select the method that looks closer to the original''.
This ensures the participants take into account the faithfulness of the reconstruction to the original image.
We select the pairs of methods to compare by performing a binary search against the ranking of previously rated images, which reduces the number of comparisons per participant.

To obtain a ranking of methods, we use the Elo~\cite{glickman1995comprehensive} rating system, widely used to rank chess players. We view each pair-wise comparison as a game between A and B, and all comparisons define a tournament.
Running Elo in this tournament (using Elo parameters $k=30$ and $N=400$), 
yields a score per method. 
Since Elo is sensitive to game order, we run a Monte Carlo simulation, shuffling the order of comparisons $N_E{=}10\,000$ times. We report the median score over the $N_E$ simulations for each method, giving us an accurate estimate of the skill rating of each method and confidence intervals. We validated that the ordering is consistent with other methods for obtaining total orderings from pairwise comparisons that do not allow the computation of confidence intervals~\cite{negahban2012iterative,coulom2008whole}.

\section{Experiments} \label{sec:evaluation}

\paragraph{Datasets} Our training set consists of a large set of high-resolution images collected from the Internet, which we downscale to a random size between 500 and 1000 pixels, and then randomly crop to $256{\times}256$. We evaluate on three diverse benchmark datasets collected independently of our training set to demonstrate that our method generalizes beyond the training distribution:
the widely used \emph{Kodak}~\cite{kodakurl} dataset (24 images), as well as the \emph{CLIC2020}~\cite{clic2020} testset (428 images), and \emph{DIV2K}~\cite{somanywaystocitediv2k} validation set (100 images). We note that we ignore color profiles of CLIC images. The latter two mostly contain high-resolution images with shorter dimension greater than 1300px (see Appendix~\ref{sec:supp:dataset} for more statistics). We emphasize that we do not adapt our models in any way to evaluate on these datasets, and that we evaluate on the full resolution images. \vsqueezehalf

\paragraph{Metrics}
We evaluate our models and the baselines in PSNR as well as the perceptual distortions LPIPS~\cite{zhang2018unreasonable} and MS-SSIM, and we use NIQE~\cite{mittal2012making}, FID~\cite{heusel2017gans}, and KID~\cite{binkowski2018demystifying} as perceptual quality indices. MS-SSIM is the most widely used perceptual metric to asses (and train) neural compression systems. 
LPIPS measures the distance in the feature space of a deep neural network originally trained for image classification, but adapted for predicting the similarity of distorted patches, 
which is validated to predict human scores for these distortions~\cite{zhang2018unreasonable}. Three network architectures were adapted in~\cite{zhang2018unreasonable}, we use the variant based on AlexNet.
NIQE is a no-reference metric, also based on assessing how strongly the statistics of a distorted image deviate from the statistics of unmodified natural images. 
In contrast to FID, non-learned features are used, and the metric entails focusing on patches selected with a ``local sharpness measure''.

FID is a widely used metrics to asses sample quality and diversity in the context of image generation, in particular for GANs. Unlike PSNR, MS-SSIM, and LPIPS, which measure the similarity between individual \emph{pairs} of images, FID assesses the similarity between the \emph{distribution} of the reference images and the distribution of the generated/distorted images. This similarity is measured in the feature space of an Inception network trained for image classification, by fitting Gaussians to the features and measuring a Wasserstein distance betweeen the Gaussians of the references and the Gaussians of the generated images. 
Heusel \etal~\cite{heusel2017gans} demonstrate that FID is consistent with increasing distortion and human perception. Furthermore, it was shown to detect common GAN failure modes such as mode dropping and mode collapse \cite{lucic2018gans}.
KID is similar, but unlike FID, is unbiased and does not make any assumptions about the distributions in the features space. 

As the distortion loss ensures global consistency of the image,
and due to the large variation of resolutions in our test sets, we calculate FID and KID on patches rather than on the full images, covering the image (details in Appendix~\ref{sec:supp:fidpatches}).
We use a patch size of 256 pixels, which yields 28\,650 patches for CLIC2020 and 6\,573 patches for DIV2K. We do not report FID or KID for Kodak, as the 24 images only yield 192 patches.  \vsqueezehalf

\paragraph{Model Variants and Baselines} 
We call our main model \emph{\textbf{High Fidelity Compression (HiFiC)}}.
To see the effect of the GAN, we train a baseline with the same architecture and same distortion loss $d=\lMSE \text{MSE} + \lP d_P$, but no GAN loss, called \emph{\textbf{Baseline (no GAN)}}.
We train all models with Adam~\cite{kingmaB14} for \million{2}{0} steps,
and initialize our GAN models with weights trained for $\lR' r+d$ 
only, which speeds up experiments (compared to training GAN models from scratch) and makes them more controllable. 
Exact values of all training hyper-parameters are tabulated in Appendix~\ref{sec:supp:training}.

We use the non-autoregressive \emph{\textbf{Mean\&Scale (M\&S) Hyperprior}} model from Minnen~\etal~\cite{minnen2018joint} as a strong baseline for an MSE-trained network.
We emphasize that this model uses the same probability model $P$ as we use, and that we train it with the same schedule as our models---the main architectural difference to ours is that Minnen \etal use a shallower auto-encoder specifically tuned for MSE.
We train \eblminnen for 15 rate points, allowing us to find a reconstruction with reasonably similar bitrate to our models for all images, and we obtain models outperforming BPG with similar performance to what is reported in~\cite{minnen2018joint} (Appendix~\ref{sec:supp:nibblerperf} shows a rate-distortion plot). When comparing against BPG, we use the highest PSNR setting, \ie, no chroma subsampling and ``slow'' mode. \vsqueezehalf

\paragraph{User Study}
We compare a total of $N_M{=}9$ methods:
We use \ename models trained for $\rtarget \in \{0.14, 0.3, 0.45\}$, denoted \textbf{\enamelo}, \textbf{\enamemi}, \textbf{\enamehi}. 
For each such model, we go through all images and select the \eblminnen model (out of our 15) that produces a reconstruction using \emph{at least} as many bits as \ename for that image. 
Additionally, we use \eblmselpips trained for $\rtarget{=}0.14$,
and BPG at two operating points, namely at $1.5{\times}$ and $2{\times}$ the bitrate of \enamemi.
The resulting bitrates do not exactly match $\rtarget$ and are shown below models in Fig.~\ref{fig:userstudy}.
We asked $N_P{=}14$ participants to complete our study.
Participants rated an average of 348 pairs of methods, taking them an average of one hour, yielding a total of 4876 comparisons. 

\section{Results} \label{sec:results}
\subsection{User Study} \label{sec:results:study}
\begin{figure}
    \centering
    \includegraphics[width=1.0\textwidth]{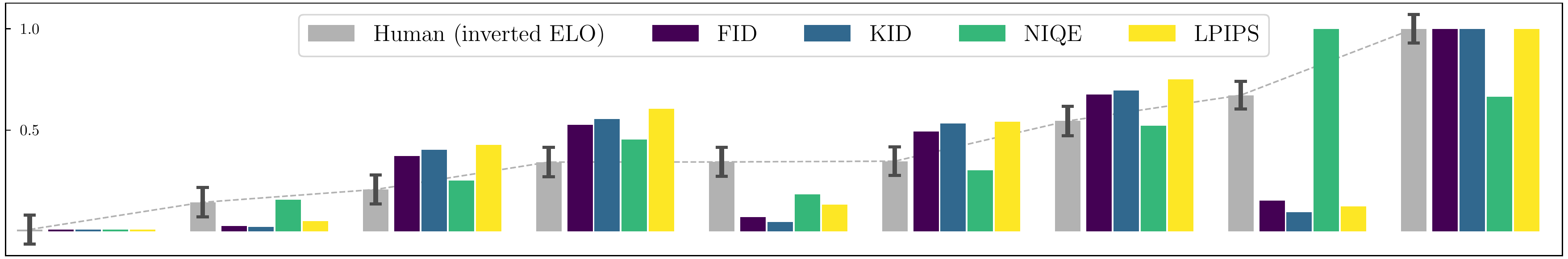}
    {\scriptsize
    \begin{tabular}{@{\hskip 1mm}p{0.111\textwidth}@{\hskip 0in}p{0.111\textwidth}@{\hskip 0in}p{0.111\textwidth}@{\hskip 0in}p{0.111\textwidth}@{\hskip 0in}p{0.111\textwidth}@{\hskip 0in}p{0.111\textwidth}@{\hskip 0in}p{0.111\textwidth}@{\hskip 0in}p{0.111\textwidth}@{\hskip 0in}p{0.111\textwidth}@{\hskip 0in}}
         \namehi \scalebox{.9}[1.0]{\emph{Ours}} & \namemi \scalebox{.9}[1.0]{\emph{Ours}} & BPG & M\&S & \namelo \scalebox{.9}[1.0]{\emph{Ours}} & BPG & M\&S & no GAN & M\&S \\
         \textbf{0.359 bpp}&\textbf{0.237 bpp}&\textbf{0.504 bpp}&\textbf{0.405 bpp}&\textbf{0.120 bpp}&\textbf{0.390 bpp}&\textbf{0.272 bpp}&\textbf{0.118 bpp}&\textbf{0.133 bpp} \\
{\fontsize{5pt}{3em}\selectfont{}MSE+LPIPS+GAN}&{\fontsize{5pt}{3em}\selectfont{}MSE+LPIPS+GAN}&&{\fontsize{5pt}{3em}\selectfont{}MSE}&{\fontsize{5pt}{3em}\selectfont{}MSE+LPIPS+GAN}&&{\fontsize{5pt}{3em}\selectfont{}MSE}&{\fontsize{5pt}{3em}\selectfont{}MSE+LPIPS}&{\fontsize{5pt}{3em}\selectfont{}MSE}\\[-2ex]
    \end{tabular}}
\caption{\label{fig:userstudy}Normalized scores for the user study, compared to perceptual metrics. 
We invert human scores such that \textbf{lower is better} for all.  Below each method, we show \emph{average} bpp, and for learned methods we show the loss components.
``no GAN'' is our baseline, using the same architecture and distortion $d$ as \enameours, but no GAN. ``M\&S'' is the \emph{Mean \& Scale Hyperprior} MSE-optimized baseline.
The study shows that training with a GAN yields reconstructions that outperform BPG at practical bitrates, for high-resolution images.
Our model at 0.237bpp is preferred to 
BPG even if BPG uses $2.1{\times}$ the bitrate, and to 
MSE optimized models even if they use $1.7{\times}$ the bitrate.
\vspace{-2.5ex}
}
\end{figure}
We visualize the outcome of the user study in Fig.~\ref{fig:userstudy}.
On the x-axis, we show the different methods sorted by the human ranking, with their average bits per pixel (bpp) on the $N_I$ images, as well as the losses used for learned methods.
We invert ELO and normalize all scores to fall between 0.01 and 1 for easier visualization.
All metrics apart from the inverted ELO are calculated on the entire images instead of user-selected crops as we want to asses the amenability of the metrics for determining ratings and these crops  would only be available through user studies.

We can observe the following: 
Our models (\name) are always preferred to MSE-based models at equal bitrates.
Comparing \enamelo to \eblmselpips, we can see that adding a GAN clearly helps for human perception.
Furthermore, \enamelo at 0.120bpp achieves similar ELO scores as BPG at 0.390bpp ($3.3{\times}$), and similar scores as \eblminnen at 0.405bpp ($3.4{\times}$). 
\enamemi at 0.237bpp is preferred to BPG when BPG uses 0.504bpp, more than $2{\times}$ the bits, and preferred to \eblminnen when it uses $1.7{\times}$ the bits. 
We note that BPG at around this bitrate is in a regime where the most severe artifacts start to disappear.
The fact that \enamemi is preferred to BPG at half the bits shows how using a GAN for neural compression can yield high fidelity images with great bit savings compared to other approaches.

Furthermore, if we fix the architecture to ours and the distortion loss to $d=\lMSE \text{MSE} + \lP d_P$, the perceptual quality indices properly rank the resulting four models (\enamelo, \enamemi, \enamehi, \eblmselpips).
However, none of the metrics would have predicted the overall human ranking. Especially FID and KID overly punish MSE-based methods, and LPIPS improves when optimized for.
On the positive side, this indicates these metrics can be used to order methods of similar architecture and distortion losses. However, we also see that currently there is no metric available to fully replace a user study.

In Appendix~\ref{sec:supp:userstudy}, we show that running the Elo tournament per user (inter-participant agreement), and per image (participant consistency at image level) yields the same trends. 

\subsection{Visual Results} \label{sec:visuals}
\paragraph{\name}
In Fig.~\ref{fig:frontpic}, we compare \enamelo 
to BPG at the same and at $2{\times}$ the bitrate. 
The crops highlight how \ename reconstructs both the hair and sweater realistically, and very close to the input. BPG at the same bitrate exhibits significant blocking and blurring artifacts, which are lessened but still present at $2{\times}$ the rate. 
In the background, we see that our full reconstruction is very similar to the original, including the skin and hat texture. 
In Appendix~\ref{sec:supp:morevisuals}, we show images from all of our datasets and compare to more methods, at various bitrates. There, we also provide download links to all reconstructions. For more visuals, see \href{https://hific.github.io}{\texttt{hific.github.io}}.

\paragraph{Failure Cases}
In Appendix~\ref{sec:supp:morevisuals}, we can also see examples of the two failure cases of \ename.
The first is very small scale text, shown in CLIC2020/25bf4, which looks typeset in another script. The second is faces that are small relative to the image, as in Kodak/14, where our \enamelo model shows high-frequency noise.

\paragraph{Previous Work using GANs} In Appendix~\ref{sec:previouswork}, we compare qualitatively to previous work by Agustsson \etal~\cite{agustsson2019extreme} and Rippel \etal~\cite{rippel17a}.

\subsection{Quantitative Results} \label{sec:results:quant}

\begin{figure}[t]
    \centering
    \vspace{-2ex}
    \includegraphics[width=\textwidth]{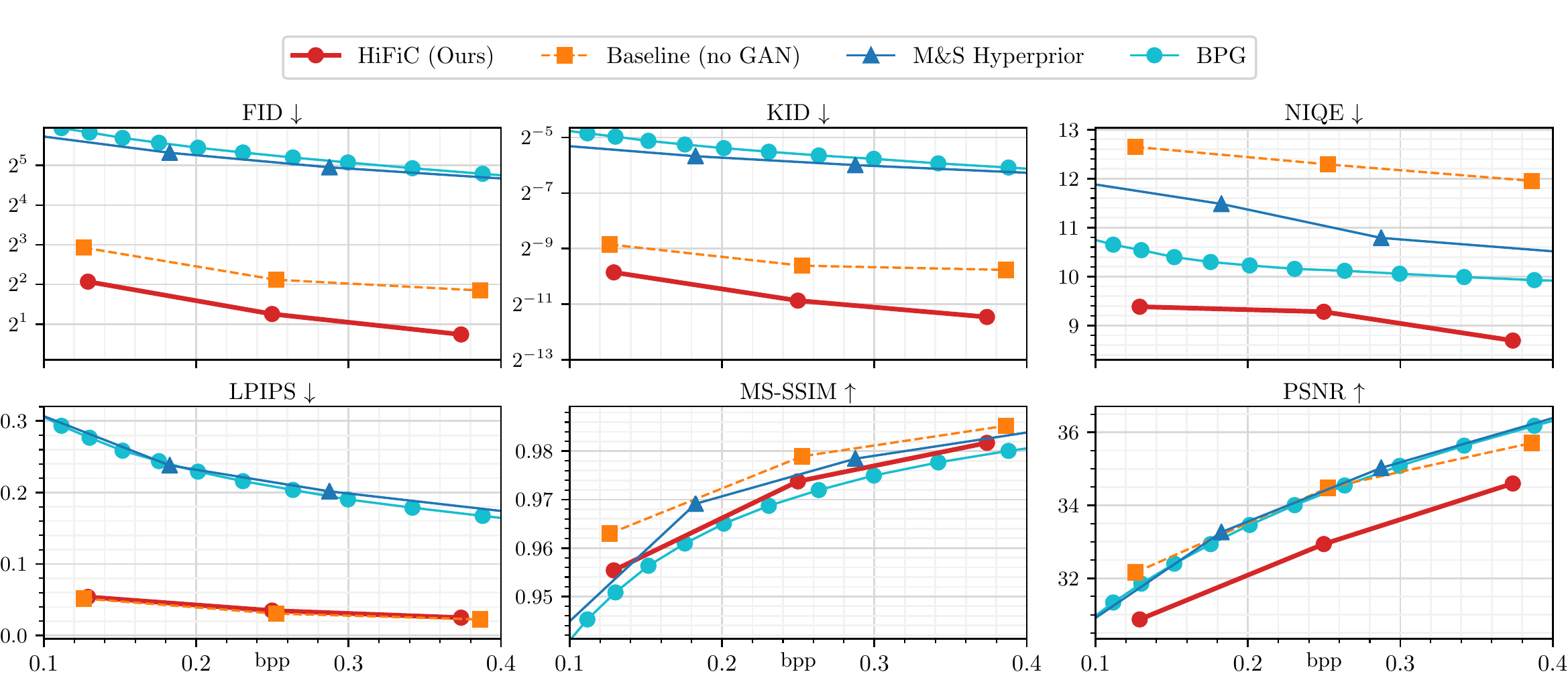}
    \vspace{-4.ex}
    \caption{Rate-distortion and -perception curves on CLIC2020. Arrows in the title indicate whether lower is better ($\downarrow$), or higher is better ($\uparrow$). Methods are described in Section~\ref{sec:evaluation}.} 
    \vspace{-3.5ex}
    \label{fig:rd_rp}
\end{figure}

\begin{wrapfigure}[12]{r}{0.44\textwidth}
\vspace{-1.8ex}
\centering
\footnotesize
\begin{overpic}[width=\linewidth,trim=-12 -10 0 10,clip]{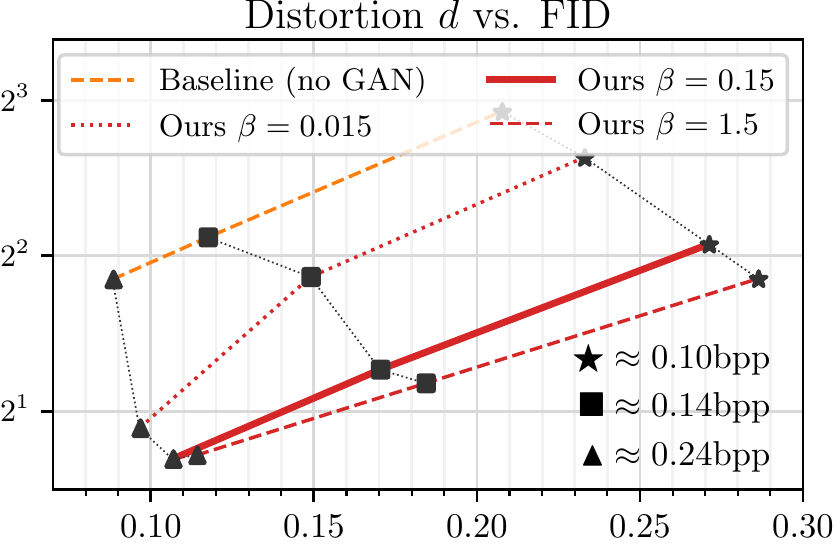}
\put (0,33) {\rotatebox{90}{FID}}
\put (40,0) {Distortion $d$}
\end{overpic}
    \caption{\label{fig:cp}Distortion-perception trade-off.}
\end{wrapfigure}
\paragraph{Effect of GAN}
In Fig.~\ref{fig:rd_rp}, we show rate-distortion and rate-perception curves using our six metrics (see Section~\ref{sec:evaluation}), on CLIC2020.
We compare \enameours, \eblmselpips, \eblminnen, and BPG. We can see that, as expected, our GAN model dominates in all perceptual quality indices, but has comparatively poor PSNR and MS-SSIM. 
Comparing \eblmselpips to \ename, which both have the same distortion  $d=\lMSE\text{MSE}+\lP d_P$, reveals that the effect of adding 
a GAN loss is consistent with the theory, and with the user study: all perceptual metrics improve,
and both components of $d$ and thus the total distortion $d$ get worse (see also next paragraph).
We observe that MSE (PSNR) is more at odds with the GAN loss than LPIPS, which gets only marginally worse when we add it.
These observations motivate us to use FID in ablation studies, as long as $d$ and the overall training setup is fixed. 
We show similar figures for the other datasets in Appendix~\ref{sec:supp:furthereval}. \vsqueezehalf

\paragraph{Distortion-Perception Trade-off}
In Fig.~\ref{fig:cp}, we show how varying the GAN loss weight $\lD$ navigates the distortion-perception trade-off. We plot FID on the y axis, and the full $d=\lMSE \text{MSE} + \lP d_P$ on the x axis.
We show \ename on a exponential grid $\lD \in \{0.015, 0.15, 1.5\}$, and \eblmselpips as a reference.
Each model is shown at three different bitrates, and models with similar bitrate are connected with black dotted lines. We see that across rates, as we increase $\lD$, the perceptual quality index improves, while $d$ suffers. This effect is lessened at higher rates (towards left of figure), possibly due to different relative loss magnitudes at this rate, \ie, $d$ is much smaller, $r$ is larger.\vsqueezehalf

\begin{figure}[b]
\centering
\begin{subfigure}[b]{0.3264\textwidth}
    \includegraphics[width=\textwidth]{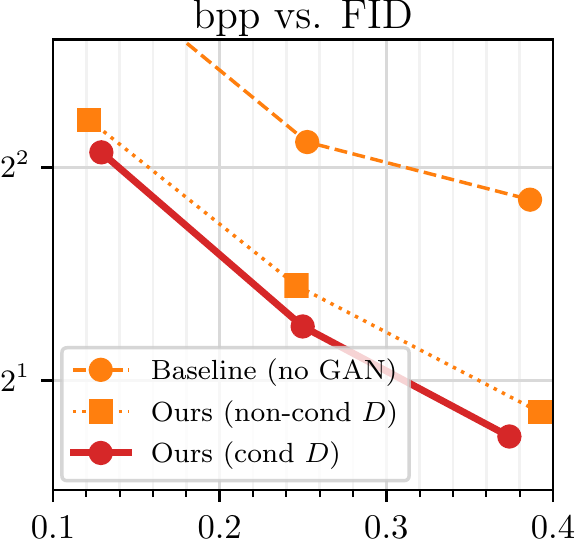}
    \vspace{-1.3ex}
    \caption{Non-conditional $D$\label{fig:condD}}
\end{subfigure}
\qquad
\begin{subfigure}[b]{0.612\textwidth}
    \includegraphics[width=\textwidth]{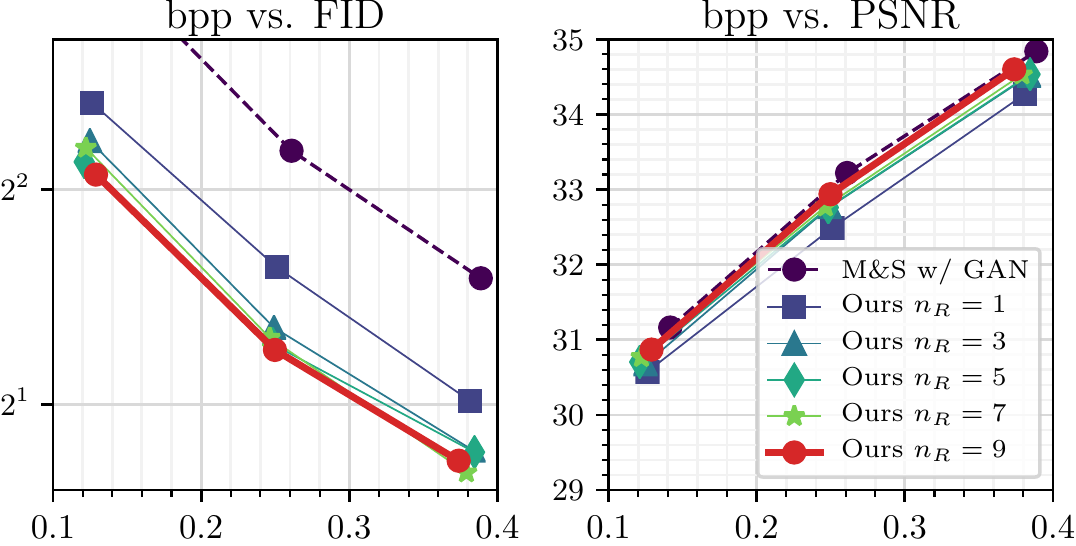}
    \vspace{-1.3ex}
    \caption{\label{fig:num_res}Varying generator capacity}
\end{subfigure}
\normalsize
    \caption{
    a) Shows the effect of a non-conditional $D$, b) shows models with different number of residual blocks $n_R$ in $G$.}
\end{figure}

\subsection{Studies} \label{sec:results:abl}
\paragraph{Discriminator: On Conditioning and Normalization}
Recall that we use a conditional $D$, in contrast to previous work. 
While our formulation can be adapted to include a non-conditional $D$, this changes training dynamics significantly: Theoretically, ignoring the distortion $d$ for a moment, $G$ can then learn to produce a natural image that is completely unrelated do with the input, and still achieve a good discriminator loss. 
While we guide $G$ with $d$, we nevertheless found that training a \emph{non-conditional} discriminator leads to images that are less sharp. The resulting models also obtain a worse FID, see Fig.~\ref{fig:condD}. \vsqueezehalf

In~\cite{wang2018highres}, InstanceNorm is used in $D$, which causes artifacts in our setup, prompting us to replace it with SpectralNorm~\cite{miyato2018spectral}. In pure generation, SpectralNorm was shown to reduce the variance of FID across multiple runs~\cite{kurach2019large}.
To explore this in our setup, we run each of the models shown in Table~\ref{tab:var} four times, and measure the mean and largest difference between runs, on our metrics. We find that in the \emph{non-conditional} setting, SpectralNorm reduces the variability of FID and KID significantly, and, to a lesser extent, also that of LPIPS and NIQE. This effect is weakened when using the \emph{conditional} $D$, where SpectralNorm only marginally reduces variability in all metrics, which we view as a further benefit of conditioning. Overall, keeping the normalization dimension fixed, we also see that conditioning $D$ improves all metrics. \vsqueezehalf

\begin{table}
\small
\centering
\begin{tabular}{lcllllllll}\toprule
&&\multicolumn{2}{l}{FID}&\multicolumn{2}{l}{KID}&\multicolumn{2}{l}{LPIPS}&\multicolumn{2}{l}{NIQE}\\[-2pt]
Norm in $D$ & Conditional $D$ & $\mu$ & $\Delta$ & $\mu$ & $\Delta$ & $\mu$ & $\Delta$ & $\mu$ & $\Delta$\\[-2pt]
\midrule
InstanceNorm &  & $5.8$ & $3.4$ & $1.1\textsc{e}^{-3}$ & $9.7\textsc{e}^{-4}$ & $6.6\textsc{e}^{-2}$ & $2.7\textsc{e}^{-3}$ & $9.0$ & $0.66$\\
SpectralNorm &  & $4.7$ & $0.14$ & $1.3\textsc{e}^{-3}$ & $1.6\textsc{e}^{-4}$ & $5.6\textsc{e}^{-2}$ & $1.2\textsc{e}^{-3}$ & $9.4$ & $0.43$\\
InstanceNorm & \checkmark & \markgray{$4.3$} & $0.68$ & \markgray{$7.1\textsc{e}^{-4}$} & $2.4\textsc{e}^{-4}$ & $6.3\textsc{e}^{-2}$ & $1.6\textsc{e}^{-3}$ & \markgray{$8.8$} & $0.64$\\
SpectralNorm & \checkmark & \markgray{$4.3$} & $0.36$ & $1.2\textsc{e}^{-3}$ & $2.2\textsc{e}^{-4}$ & \markgray{$5.5\textsc{e}^{-2}$} & $7.4\textsc{e}^{-4}$ & $9.4$ & $0.32$\\[-2pt]
\bottomrule
\end{tabular}
\vspace{2pt}
\caption{\label{tab:var} Exploring across-run variation. Each model (row) is run four times in exactly the same configuration, and we show mean $\mu$ and difference between maximal and minimal value $\Delta$.}
\end{table}

\newpage

\paragraph{Instance Norm in Auto-Encoder}
\begin{wrapfigure}[5]{r}{0.2\textwidth}
\captionsetup{labelformat=empty,singlelinecheck=false,skip=1pt,font=small}
\captionsetup{justification=raggedright}
\raggedright
\vspace{-0.6em}
    \includegraphics[width=0.66\linewidth]{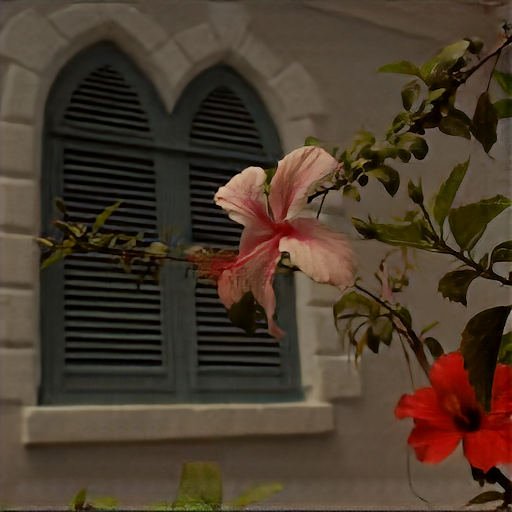}\hfill
    \includegraphics[width=0.33\linewidth]{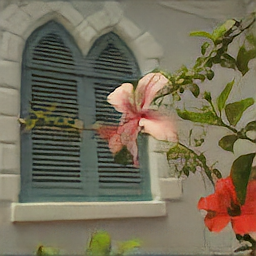}\\
  \caption{\tiny $512{\times}512$\hfill $256{\times}256$}
\end{wrapfigure}
We visualize the darkening caused by InstanceNorm in $E, G$ (see Section~\ref{sec:arch}) in the inset figure, where a model is evaluated on an image at a resolution of $512{\times}512$px as well as at the training resolution ($256{\times}256$px).
We also explored using BatchNorm~\cite{ioffe2015batch} but found that this resulted in unstable training in our setup.

\paragraph{Generator Capacity}
By default, we use $n_R{=}9$ residual blocks in $G$ (see Fig.~\ref{fig:arch}). We show the result of varying $n_R$ in Fig.~\ref{fig:num_res}. While $n_R{=}1$ gives significantly worse FID and PSNR, both metrics start to saturate around $n_R{=}5$. We also show the \eblminnen baseline trained with our loss (``M\&S w/ GAN''), \ie, with LPIPS and a conditional $D$, and exactly the same training schedule. While this yields slightly better PSNR as \name, FID drops by a factor 2, further indicating the importance of capacity, but also how suited the hyperprior architecture is for MSE training.  \vsqueezehalf

\begin{figure}[t]
\tiny
\centering
\captionsetup[subfigure]{skip=1pt,singlelinecheck=false}
\begin{subfigure}{0.33\textwidth}
\includegraphics[width=\textwidth]{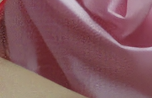}
\caption{\label{fig:perc:no}no $d_P$}
\end{subfigure}
\begin{subfigure}{0.33\textwidth}
\includegraphics[width=\textwidth]{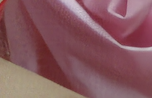}
\caption{\label{fig:perc:vgg}$d_P=\text{VGG}$}
\end{subfigure}
\begin{subfigure}{0.33\textwidth}
\includegraphics[width=\textwidth]{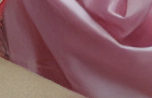}
\caption{\label{fig:perc:lpips}$d_P=\text{LPIPS}$}
\end{subfigure}
\vspace{-2.3ex}
\caption{\label{fig:perc}Effect of varying the perceptual distortion $d_P$. All models were also trained with an MSE loss and a GAN loss.
}
\end{figure}

\paragraph{Training Stability and Losses}
\begin{wrapfigure}{r}{0.35\textwidth}
\tiny
\centering
\begin{overpic}[width=0.49\linewidth]{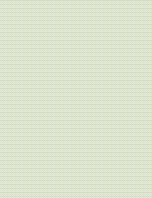}
\put (10,5) {\scriptsize \textbf{Learning $\bm E$}}
\end{overpic}
\begin{overpic}[width=0.49\linewidth]{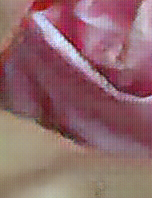}
\put (10,5) {\scriptsize \textbf{\textcolor{nicegray}{Frozen $\bm E$}}}
\end{overpic}
\caption{\label{fig:perc:onlygan}Training with a GAN loss, but no MSE or $d_P$.} 
\end{wrapfigure}
Training GANs for unconditional or class-conditional image generation is notoriously hard \cite{lucic2018gans, miyato2018spectral, brock2018large}. However, our setup is inherently different through the distortion loss $d$, which guides the optimization at the pixel-level. 
While our initialization scheme of using weights trained to minimize $\lambda'r + d$ (see Section~\ref{sec:evaluation}) can help, MSE seems to be the driving force for stability in our setup: 
For Fig.~\ref{fig:perc:onlygan}, we initialize the network with $\lambda'r + d$-optimized weights, then train with a GAN loss but without a distortion loss $d$. We compare fixing $E$ to using the $\lambda'r + d$ weights, to jointly learning $E$. We see that the GAN collapses if we learn $E$.
Additionally, we explore varying $d_P$: In Fig.~\ref{fig:perc:no}, we see that removing $d_P$ introduces significant gridding artifacts. Using $d_P{=}\text{VGG}$, \ie, $L_1$ in the feature space of VGG, some gridding is still visible (Fig.~\ref{fig:perc:vgg}). Using $d_P{=}\text{LPIPS}$ alleviates these artifacts. We note that only using LPIPS \emph{without a GAN} also produces minor repetitive patterns as well as less sharp reconstructions, which is validated by the fact that \eblmselpips ranks worse than our GAN models in the user study. We show a visual example in Appendix~\ref{sec:suppl:lpipsonly}.

\section{Conclusion}

In this paper, we showed how optimizing a neural compression scheme with a GAN yields reconstructions with high perceptual fidelity that are visually close to the input, and that are preferred to previous methods even when these approaches use more than double the bits.
We evaluated our approach with a diverse set of metrics and interpreted the results with rate-distortion-perception theory.
Guided by the metrics, we studied various components of our architecture and loss. 
By comparing our metrics with the outcome of the user study, we showed that no existing metric is perfect for ordering arbitrary models right now, but that using FID and KID can be a valuable tool in exploring architectures and other design choices. 
Future work could focus on further studying perceptual score indices and metrics (to better predict human preferences), and further investigate failure cases such as small faces.
Additionally, generative video compression is a very interesting direction for future work. To ensure temporal consistency, mechanisms to enforce smoothness could be borrowed from, e.g., GAN-based video super resolution models.

\newpage

\section*{Broader Impact}
Users of our compression method can benefit from better reconstructions at lower bitrates, reducing the amount of storage needed to save pictures and the bandwidth required to transmit pictures. The latter is important as wireless technology typically lags behind user trends which have been continuously requiring higher bandwidths over the past decades, and there is no end in sight with emerging applications such as virtual reality. Furthermore, better compression technology improves accessibility in developing areas of the world where the wireless infrastructure is less performant and robust than in developed countries. It is important to keep in mind that we employ a generator $G$ that in theory can produce images that are very different from the input. While this is the case for \emph{any} lossy image compression algorithm, this has a bigger impact here as we specifically train $G$ for realistic reconstructions. Therefore, we emphasize that our method is not suitable for sensitive image contents, such as, \eg, storing medical images, or important documents.

\section*{Funding}

This work was done at Google Research.

\section*{Acknowledgments}

The authors would like to thank Johannes Balle, Sergi Caelles, Sander Dielmann, and David Minnen for the insightful discussions and feedback.

{\footnotesize
\bibliographystyle{plain}
\bibliography{ref}

\begin{thebibliography}{10}

\bibitem{agustsson2017soft}
Eirikur Agustsson, Fabian Mentzer, Michael Tschannen, Lukas Cavigelli, Radu
  Timofte, Luca Benini, and Luc~V Gool.
\newblock Soft-to-hard vector quantization for end-to-end learning compressible
  representations.
\newblock In {\em Advances in Neural Information Processing Systems}, pages
  1141--1151, 2017.

\bibitem{somanywaystocitediv2k}
Eirikur Agustsson and Radu Timofte.
\newblock Ntire 2017 challenge on single image super-resolution: Dataset and
  study.
\newblock In {\em The IEEE Conference on Computer Vision and Pattern
  Recognition (CVPR) Workshops}, July 2017.

\bibitem{agustsson2019extreme}
Eirikur Agustsson, Michael Tschannen, Fabian Mentzer, Radu Timofte, and Luc~Van
  Gool.
\newblock Generative adversarial networks for extreme learned image
  compression.
\newblock In {\em The IEEE International Conference on Computer Vision (ICCV)},
  October 2019.

\bibitem{ba2016layer}
Jimmy~Lei Ba, Jamie~Ryan Kiros, and Geoffrey~E Hinton.
\newblock Layer normalization.
\newblock {\em arXiv preprint arXiv:1607.06450}, 2016.

\bibitem{balle2016end}
Johannes Ball{\'e}, Valero Laparra, and Eero~P Simoncelli.
\newblock End-to-end optimized image compression.
\newblock {\em arXiv preprint arXiv:1611.01704}, 2016.

\bibitem{balle2018variational}
Johannes Ball{\'e}, David Minnen, Saurabh Singh, Sung~Jin Hwang, and Nick
  Johnston.
\newblock Variational image compression with a scale hyperprior.
\newblock In {\em International Conference on Learning Representations (ICLR)},
  2018.

\bibitem{bpgurl}
Fabrice Bellard.
\newblock {BPG Image format}.
\newblock \url{https://bellard.org/bpg/}.

\bibitem{binkowski2018demystifying}
Miko{\l}aj Bi{\'n}kowski, Dougal~J Sutherland, Michael Arbel, and Arthur
  Gretton.
\newblock Demystifying mmd gans.
\newblock In {\em International Conference on Learning Representations}, 2018.

\bibitem{blau2019rethinking}
Yochai Blau and Tomer Michaeli.
\newblock Rethinking lossy compression: The rate-distortion-perception
  tradeoff.
\newblock {\em arXiv preprint arXiv:1901.07821}, 2019.

\bibitem{brock2018large}
Andrew Brock, Jeff Donahue, and Karen Simonyan.
\newblock Large scale gan training for high fidelity natural image synthesis.
\newblock In {\em International Conference on Learning Representations}, 2019.

\bibitem{chen2018image}
Jingwen Chen, Jiawei Chen, Hongyang Chao, and Ming Yang.
\newblock Image blind denoising with generative adversarial network based noise
  modeling.
\newblock In {\em The IEEE Conference on Computer Vision and Pattern
  Recognition (CVPR)}, June 2018.

\bibitem{coulom2008whole}
R{\'e}mi Coulom.
\newblock Whole-history rating: A bayesian rating system for players of
  time-varying strength.
\newblock In {\em International Conference on Computers and Games}, pages
  113--124. Springer, 2008.

\bibitem{coulombe2009low}
St{\'e}phane Coulombe and Steven Pigeon.
\newblock Low-complexity transcoding of jpeg images with near-optimal quality
  using a predictive quality factor and scaling parameters.
\newblock {\em IEEE Transactions on Image Processing}, 19(3):712--721, 2009.

\bibitem{cover2012elements}
Thomas~M Cover and Joy~A Thomas.
\newblock {\em Elements of information theory}.
\newblock John Wiley \& Sons, 2012.

\bibitem{galteri2017deep}
Leonardo Galteri, Lorenzo Seidenari, Marco Bertini, and Alberto Del~Bimbo.
\newblock Deep generative adversarial compression artifact removal.
\newblock In {\em Proceedings of the IEEE Conference on Computer Vision and
  Pattern Recognition}, pages 4826--4835, 2017.

\bibitem{glickman1995comprehensive}
Mark~E Glickman.
\newblock A comprehensive guide to chess ratings.
\newblock {\em American Chess Journal}, 3(1):59--102, 1995.

\bibitem{goodfellow2014generative}
Ian Goodfellow, Jean Pouget-Abadie, Mehdi Mirza, Bing Xu, David Warde-Farley,
  Sherjil Ozair, Aaron Courville, and Yoshua Bengio.
\newblock Generative adversarial nets.
\newblock In {\em Advances in neural information processing systems}, pages
  2672--2680, 2014.

\bibitem{heusel2017gans}
Martin Heusel, Hubert Ramsauer, Thomas Unterthiner, Bernhard Nessler, and Sepp
  Hochreiter.
\newblock Gans trained by a two time-scale update rule converge to a local nash
  equilibrium.
\newblock In {\em Advances in Neural Information Processing Systems}, pages
  6626--6637, 2017.

\bibitem{ioffe2015batch}
Sergey Ioffe and Christian Szegedy.
\newblock Batch normalization: Accelerating deep network training by reducing
  internal covariate shift.
\newblock In {\em International Conference on Machine Learning}, pages
  448--456, 2015.

\bibitem{karras2017progressive}
Tero Karras, Timo Aila, Samuli Laine, and Jaakko Lehtinen.
\newblock Progressive growing of gans for improved quality, stability, and
  variation.
\newblock In {\em International Conference on Learning Representations (ICLR)},
  2017.

\bibitem{karras2019style}
Tero Karras, Samuli Laine, and Timo Aila.
\newblock A style-based generator architecture for generative adversarial
  networks.
\newblock In {\em Proceedings of the IEEE Conference on Computer Vision and
  Pattern Recognition}, pages 4401--4410, 2019.

\bibitem{kingmaB14}
Diederik~P. Kingma and Jimmy Ba.
\newblock Adam: {A} method for stochastic optimization.
\newblock {\em CoRR}, abs/1412.6980, 2014.

\bibitem{kodakurl}
{Kodak PhotoCD dataset}.
\newblock \url{http://r0k.us/graphics/kodak/}.

\bibitem{kurach2019large}
Karol Kurach, Mario Lu{\v{c}}i{\'c}, Xiaohua Zhai, Marcin Michalski, and
  Sylvain Gelly.
\newblock A large-scale study on regularization and normalization in gans.
\newblock In {\em International Conference on Machine Learning}, pages
  3581--3590, 2019.

\bibitem{ledig2017photo}
Christian Ledig, Lucas Theis, Ferenc Huszar, Jose Caballero, Andrew Cunningham,
  Alejandro Acosta, Andrew Aitken, Alykhan Tejani, Johannes Totz, Zehan Wang,
  et~al.
\newblock Photo-realistic single image super-resolution using a generative
  adversarial network.
\newblock In {\em Proceedings of the IEEE Conference on Computer Vision and
  Pattern Recognition}, pages 4681--4690, 2017.

\bibitem{lee2018context}
Jooyoung Lee, Seunghyun Cho, and Seung-Kwon Beack.
\newblock Context-adaptive entropy model for end-to-end optimized image
  compression.
\newblock {\em arXiv preprint arXiv:1809.10452}, 2018.

\bibitem{li2019positionalnorm}
Boyi Li, Felix Wu, Kilian~Q Weinberger, and Serge Belongie.
\newblock Positional normalization.
\newblock In H.~Wallach, H.~Larochelle, A.~Beygelzimer, F.~d\textquotesingle
  Alch\'{e}-Buc, E.~Fox, and R.~Garnett, editors, {\em Advances in Neural
  Information Processing Systems 32}, pages 1622--1634. Curran Associates,
  Inc., 2019.

\bibitem{li2017learning}
Mu~Li, Wangmeng Zuo, Shuhang Gu, Debin Zhao, and David Zhang.
\newblock Learning convolutional networks for content-weighted image
  compression.
\newblock {\em arXiv preprint arXiv:1703.10553}, 2017.

\bibitem{lucic2018gans}
Mario Lucic, Karol Kurach, Marcin Michalski, Sylvain Gelly, and Olivier
  Bousquet.
\newblock Are gans created equal? a large-scale study.
\newblock In {\em Advances in Neural Information Processing Systems}, pages
  700--709, 2018.

\bibitem{marpe2003context}
Detlev Marpe, Heiko Schwarz, and Thomas Wiegand.
\newblock Context-based adaptive binary arithmetic coding in the h. 264/avc
  video compression standard.
\newblock {\em IEEE Transactions on circuits and systems for video technology},
  13(7):620--636, 2003.

\bibitem{mentzer2018conditional}
Fabian Mentzer, Eirikur Agustsson, Michael Tschannen, Radu Timofte, and Luc
  Van~Gool.
\newblock Conditional probability models for deep image compression.
\newblock In {\em IEEE Conference on Computer Vision and Pattern Recognition
  (CVPR)}, 2018.

\bibitem{minnen2018joint}
David Minnen, Johannes Ball{\'e}, and George~D Toderici.
\newblock Joint autoregressive and hierarchical priors for learned image
  compression.
\newblock In {\em Advances in Neural Information Processing Systems}, pages
  10771--10780, 2018.

\bibitem{minnen2020channel}
David Minnen and Saurabh Singh.
\newblock Channel-wise autoregressive entropy models for learned image
  compression.
\newblock {\em arXiv preprint arXiv:2007.08739}, 2020.

\bibitem{mirza2014conditional}
Mehdi Mirza and Simon Osindero.
\newblock Conditional generative adversarial nets.
\newblock {\em arXiv preprint arXiv:1411.1784}, 2014.

\bibitem{mittal2012making}
Anish Mittal, Rajiv Soundararajan, and Alan~C Bovik.
\newblock Making a “completely blind” image quality analyzer.
\newblock {\em IEEE Signal Processing Letters}, 20(3):209--212, 2012.

\bibitem{miyato2018spectral}
Takeru Miyato, Toshiki Kataoka, Masanori Koyama, and Yuichi Yoshida.
\newblock Spectral normalization for generative adversarial networks.
\newblock In {\em International Conference on Learning Representations}, 2018.

\bibitem{negahban2012iterative}
Sahand Negahban, Sewoong Oh, and Devavrat Shah.
\newblock Iterative ranking from pair-wise comparisons.
\newblock In {\em Advances in neural information processing systems}, pages
  2474--2482, 2012.

\bibitem{park2019semantic}
Taesung Park, Ming-Yu Liu, Ting-Chun Wang, and Jun-Yan Zhu.
\newblock Semantic image synthesis with spatially-adaptive normalization.
\newblock In {\em Proceedings of the IEEE Conference on Computer Vision and
  Pattern Recognition}, pages 2337--2346, 2019.

\bibitem{rippel17a}
Oren Rippel and Lubomir Bourdev.
\newblock Real-time adaptive image compression.
\newblock In {\em Proceedings of the 34th International Conference on Machine
  Learning}, volume~70 of {\em Proceedings of Machine Learning Research}, pages
  2922--2930, International Convention Centre, Sydney, Australia, 06--11 Aug
  2017. PMLR.

\bibitem{santurkar2017generative}
Shibani Santurkar, David Budden, and Nir Shavit.
\newblock Generative compression.
\newblock {\em arXiv preprint arXiv:1703.01467}, 2017.

\bibitem{simonyan2014very}
Karen Simonyan and Andrew Zisserman.
\newblock Very deep convolutional networks for large-scale image recognition.
\newblock {\em arXiv preprint arXiv:1409.1556}, 2014.

\bibitem{sullivan2012overview}
Gary~J Sullivan, Jens-Rainer Ohm, Woo-Jin Han, and Thomas Wiegand.
\newblock Overview of the high efficiency video coding (hevc) standard.
\newblock {\em IEEE Transactions on circuits and systems for video technology},
  22(12):1649--1668, 2012.

\bibitem{jpeg2000taubman}
David~S. Taubman and Michael~W. Marcellin.
\newblock {\em JPEG 2000: Image Compression Fundamentals, Standards and
  Practice}.
\newblock Kluwer Academic Publishers, Norwell, MA, USA, 2001.

\bibitem{theis2017lossy}
Lucas Theis, Wenzhe Shi, Andrew Cunningham, and Ferenc Huszar.
\newblock Lossy image compression with compressive autoencoders.
\newblock In {\em International Conference on Learning Representations (ICLR)},
  2017.

\bibitem{toderici2015variable}
George Toderici, Sean~M O'Malley, Sung~Jin Hwang, Damien Vincent, David Minnen,
  Shumeet Baluja, Michele Covell, and Rahul Sukthankar.
\newblock Variable rate image compression with recurrent neural networks.
\newblock {\em arXiv preprint arXiv:1511.06085}, 2015.

\bibitem{clic2020}
George Toderici, Lucas Theis, Nick Johnston, Eirikur Agustsson, Fabian Mentzer,
  Johannes Ball{\'e}, Wenzhe Shi, and Radu Timofte.
\newblock {\em CLIC 2020: Challenge on Learned Image Compression}, 2020.
\newblock \url{http://compression.cc}.

\bibitem{toderici2016full}
George Toderici, Damien Vincent, Nick Johnston, Sung~Jin Hwang, David Minnen,
  Joel Shor, and Michele Covell.
\newblock Full resolution image compression with recurrent neural networks.
\newblock {\em arXiv preprint arXiv:1608.05148}, 2016.

\bibitem{tschannen2018deep}
Michael Tschannen, Eirikur Agustsson, and Mario Lucic.
\newblock Deep generative models for distribution-preserving lossy compression.
\newblock In {\em Advances in Neural Information Processing Systems}, pages
  5929--5940, 2018.

\bibitem{ulyanov2016instance}
Dmitry Ulyanov, Andrea Vedaldi, and Victor Lempitsky.
\newblock Instance normalization: The missing ingredient for fast stylization.
\newblock {\em arXiv preprint arXiv:1607.08022}, 2016.

\bibitem{jpeg1992wallace}
Gregory~K Wallace.
\newblock The {JPEG} still picture compression standard.
\newblock {\em IEEE transactions on consumer electronics}, 38(1):xviii--xxxiv,
  1992.

\bibitem{wang2018highres}
Ting-Chun Wang, Ming-Yu Liu, Jun-Yan Zhu, Andrew Tao, Jan Kautz, and Bryan
  Catanzaro.
\newblock High-resolution image synthesis and semantic manipulation with
  conditional gans.
\newblock In {\em IEEE Conference on Computer Vision and Pattern Recognition
  (CVPR)}, 2018.

\bibitem{SSIM-MS}
Z.~Wang, E.~P. Simoncelli, and A.~C. Bovik.
\newblock Multiscale structural similarity for image quality assessment.
\newblock In {\em Asilomar Conference on Signals, Systems Computers, 2003},
  volume~2, pages 1398--1402 Vol.2, Nov 2003.

\bibitem{SSIM}
Zhou Wang, A.~C. Bovik, H.~R. Sheikh, and E.~P. Simoncelli.
\newblock Image quality assessment: from error visibility to structural
  similarity.
\newblock {\em IEEE Transactions on Image Processing}, 13(4):600--612, April
  2004.

\bibitem{webpurl}
{WebP Image format}.
\newblock \url{https://developers.google.com/speed/webp/}.

\bibitem{wu2018group}
Yuxin Wu and Kaiming He.
\newblock Group normalization.
\newblock In {\em Proceedings of the European conference on computer vision
  (ECCV)}, pages 3--19, 2018.

\bibitem{xu2015empirical}
Bing Xu, Naiyan Wang, Tianqi Chen, and Mu~Li.
\newblock Empirical evaluation of rectified activations in convolutional
  network.
\newblock {\em arXiv preprint arXiv:1505.00853}, 2015.

\bibitem{zhang2018unreasonable}
Richard Zhang, Phillip Isola, Alexei~A Efros, Eli Shechtman, and Oliver Wang.
\newblock The unreasonable effectiveness of deep features as a perceptual
  metric.
\newblock In {\em Proceedings of the IEEE Conference on Computer Vision and
  Pattern Recognition}, pages 586--595, 2018.

\end{thebibliography}
}

\FloatBarrier
\newpage

\appendix

\renewcommand{\thefigure}{A\arabic{figure}}

\setcounter{figure}{0}

\section{Supplementary Material -- High-Fidelity Generative Image Compression}

We show further details of our method in ~\ref{sec:supp:nibblerperf}--\ref{sec:supp:dataset}, and more visual results in Appendix~\ref{sec:supp:morevisuals}.

%
%
\subsection{Comparing MSE models based on Minnen \etal~\texorpdfstring{\cite{minnen2018joint}}{2018}} \label{sec:supp:nibblerperf}

\begin{figure}[ht]
\vspace{-3ex}
    \centering
    \includegraphics[width=0.5\textwidth]{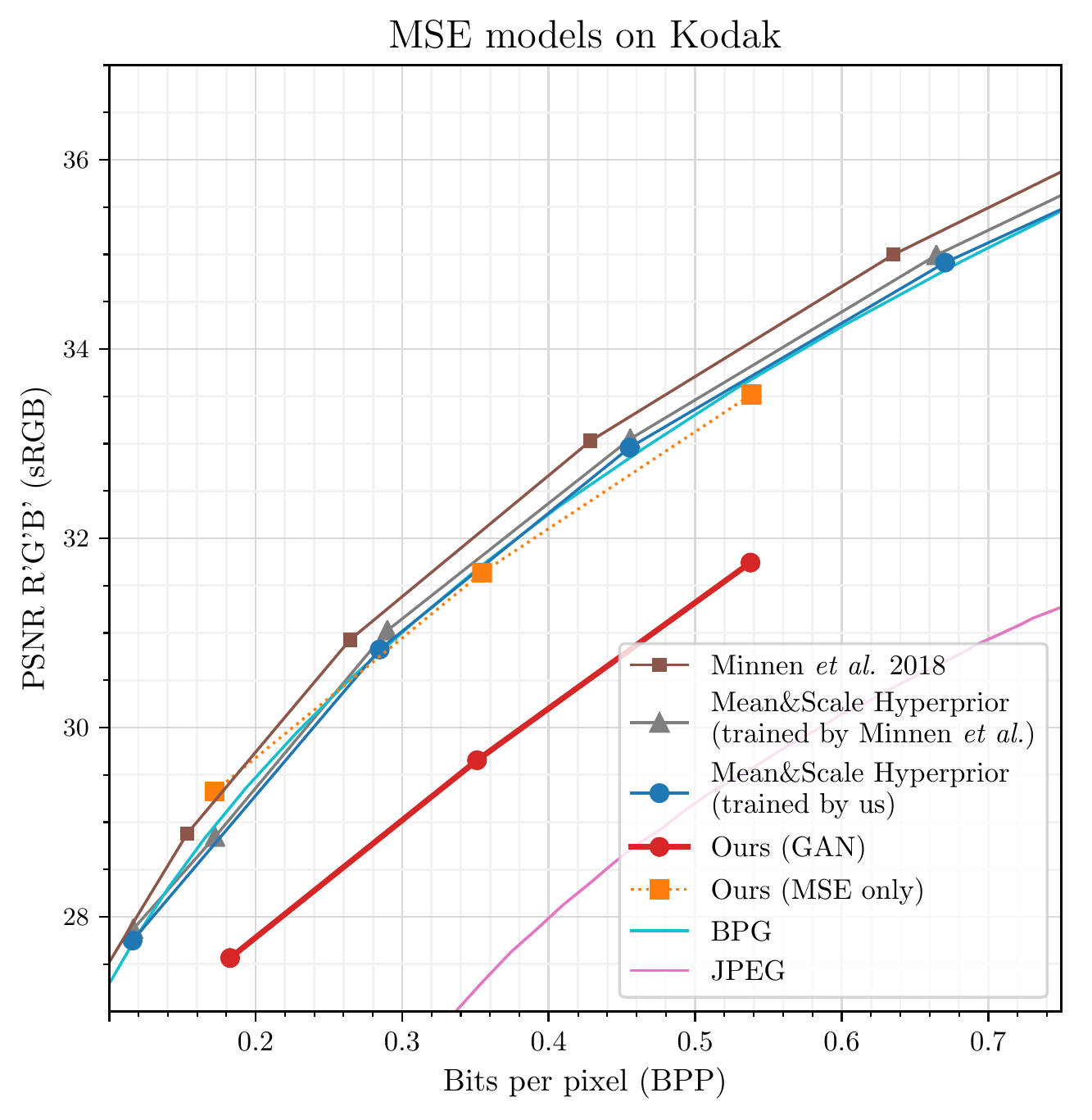}
\vspace{-2ex}
    \caption{Comparing the \emph{Mean \& Scale Hyperprior} model trained by us to the one reported by Minnen \etal~\cite{minnen2018joint}, and to their main model, denoted ``Minnen \etal\ 2018''.}
    \label{fig:nibblers}
\end{figure}

To validate that the \eblminnen model trained by us is a strong MSE model,
we compare it to models reported by Minnen \etal~\cite{minnen2018joint} in Fig.~\ref{fig:nibblers}. 
First, we compare the model trained by us to the ``Mean \& Scale Hyperprior'' baseline reported in~\cite{minnen2018joint}, which is equivalent to the \eblminnen model in terms of architecture. We observe a very minor drop in PSNR, which is likely attributable to our training schedule of \million{2}{0} steps, vs.\ the \million{6}{0} steps used in~\cite{minnen2018joint}. Then, we compare to the fully autoregressive main model of~\cite{minnen2018joint}, denoted ``Minnen \etal\ 2018''. We can see that this model is on average ${\approx}0.4\text{dB}$ better than our \eblminnen model. It is important to note that a) an auto-regressive probability model would also increase the performance of all of our models, and b) that in the user study, our GAN models were preferred to MSE-trained Hyperprior models even when the Hyperprior models used $4{\times}$ the bitrate, which amounts roughly to a ${\approx}4\text{dB}$ PSNR difference.

\subsection{Using LPIPS without a GAN loss} \label{sec:suppl:lpipsonly}

\begin{figure}[hb]
\captionsetup[subfigure]{labelformat=empty,singlelinecheck=false,skip=1pt,font=small}
\begin{subfigure}[b]{.33\textwidth}
\includegraphics[width=\linewidth]{%
    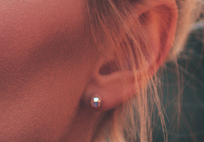}
\caption{Orignal}
\end{subfigure}
\begin{subfigure}[b]{.33\textwidth}
\includegraphics[width=\linewidth]{%
    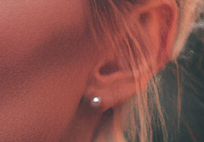}
\caption{$r$, MSE, LPIPS, GAN (\enamelo)}
\end{subfigure}
\begin{subfigure}[b]{.33\textwidth}
\includegraphics[width=\linewidth]{%
    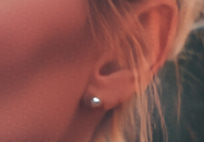}
\caption{$r$, MSE, LPIPS (\eblmselpips)}
\end{subfigure}\vspace{-1ex}
    \caption{\label{fig:lpipsonly}LPIPS without a GAN (right) leads to gridding and less sharpness than using a GAN (middle). We show the loss components below the figure, where $r$ is the rate loss. \emph{Best viewed on screen.}}
\end{figure}

We saw in the user study that the model trained for MSE and LPIPS without a GAN loss (\eblmselpips) ranks worse than the one using a GAN loss (\enamelo). In Fig.~\ref{fig:lpipsonly}, we visualize a reconstruction of \eblmselpips to show that this training setup can also cause gridding artifacts on some images. Furthermore, we see that adding a GAN loss leads to sharper reconstructions and to a faithful preservation of the image noise.

\subsection{Qualitative Comparison to Previous Generative Approaches}\label{sec:previouswork}

\begin{figure}[ht]
\captionsetup[subfigure]{labelformat=empty,singlelinecheck=false,skip=1pt,font=small}
\centering
\begin{subfigure}[t]{.35\textwidth}
  \includegraphics[trim={1.123cm 1.12875cm 1.123cm 1.12875cm},clip,width=\linewidth]{%
    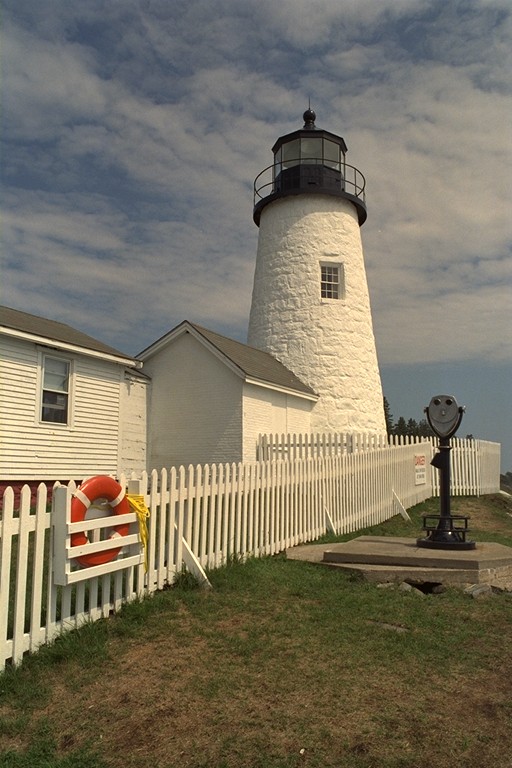}
  \caption{Original}
\end{subfigure}
\begin{subfigure}[t]{.35\textwidth}
  \includegraphics[trim={1.123cm 1.12875cm 1.123cm 1.12875cm},clip,width=\linewidth]{%
    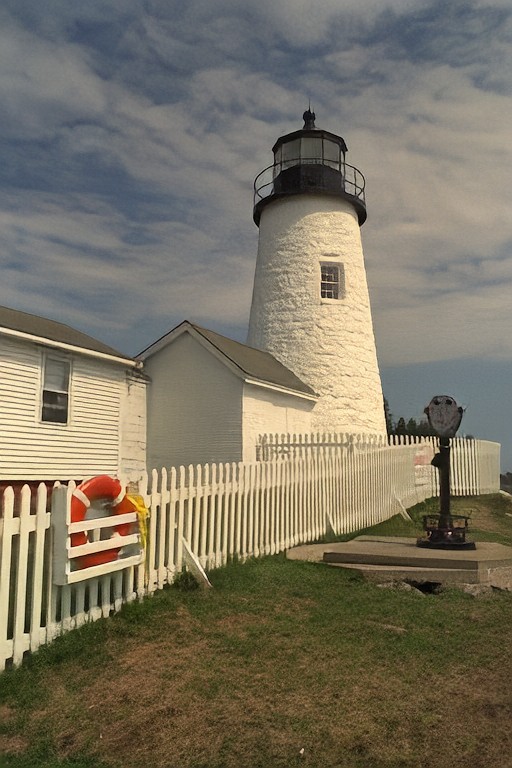}
  \caption{\nameourslo: 0.162bpp}
\end{subfigure}
\begin{subfigure}[t]{.35\textwidth}
  \includegraphics[width=\linewidth]{%
    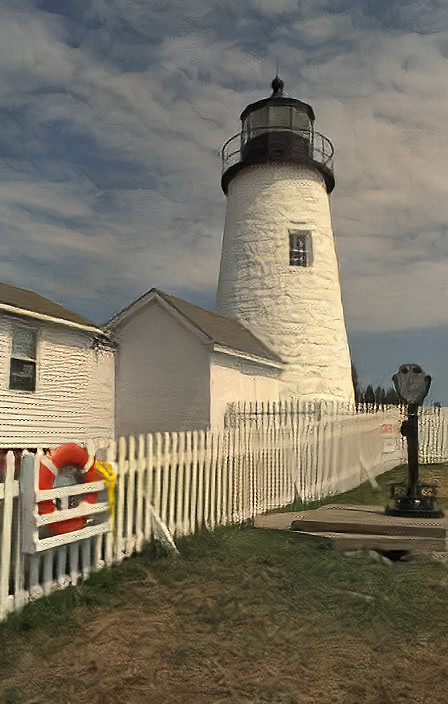}
  \caption{Rippel \etal~\cite{rippel17a}: 0.194bpp}
\end{subfigure}
\begin{subfigure}[t]{.35\textwidth}
  \includegraphics[trim={1.123cm 1.12875cm 1.123cm 1.12875cm},clip,width=\linewidth]{%
    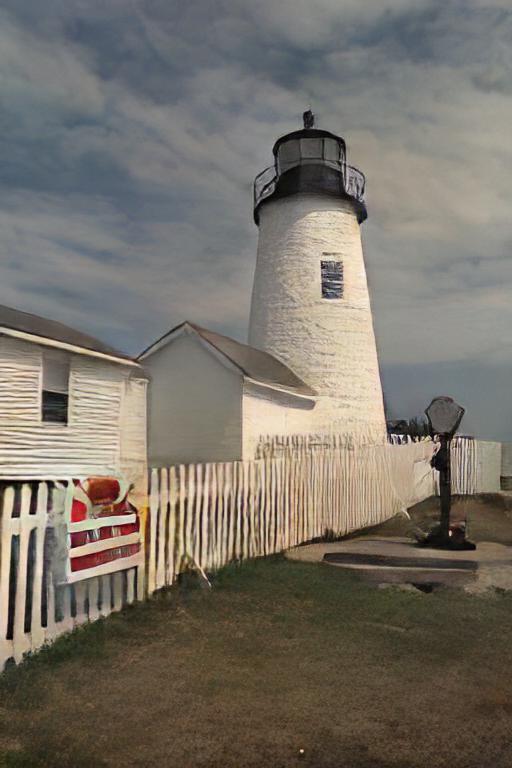}
  \caption{Agustsson \etal~\cite{agustsson2019extreme}: 0.0668bpp}
\end{subfigure}
    \caption{\label{fig:others}Comparison between neural compression approaches using an adverserial loss on Kodak/19. Note the high-frequency artifacts in the tower and fence for~\cite{rippel17a}, as well as the different grass texture. Note that the comparison to~\cite{agustsson2019extreme} is  not ideal, as their highest bpp reconstructions use less than half the rate of \ename, but we see that the reconstruction deviates significantly from the input. Furthermore, both~\cite{rippel17a} and~\cite{agustsson2019extreme} exhibit a small color shift.}
\end{figure}

We compare against Agustsson \etal~\cite{agustsson2019extreme} as well as Rippel \etal~\cite{rippel17a} in Fig.~\ref{fig:others} on an image from the Kodak dataset, as both incorporated adverserial losses into their training. We do no further comparisons, as Agustsson \etal targeted ``extremely low'' bitrates and thus operate in a very different regime than our models, and Rippel \etal only released a handful of images.

\subsection{ChannelNorm} \label{sec:channelnormdetails}

\begin{figure}
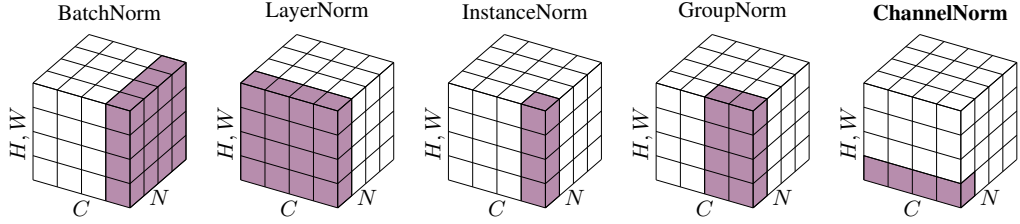

\small
\tdplotsetmaincoords{60}{30}  
\normcube{BatchNorm}{(3,0,0) -- (3,0,4) -- (3,4,4) -- (4,4,4) -- (4,4,0) -- (4,0,0)}%
\quad %
\normcube{LayerNorm}{(0,0,0) -- (0,0,4) -- (0,1,4) -- (4,1,4) -- (4,1,0) -- (4,0,0)}%
\quad %
\normcube{InstanceNorm}{(3,0,0) -- (3,0,4) -- (3,1,4) -- (4,1,4) -- (4,1,0) -- (4,0,0)}%
\quad %
\normcube{GroupNorm}{(2,0,0) -- (2,0,4) -- (2,1,4) -- (4,1,4) -- (4,1,0) -- (4,0,0)}%
\quad %
\normcube{\textbf{ChannelNorm}}{(0,0,0) -- (0,0,1) -- (4,0,1) -- (4,1,1) -- (4,1,0) -- (4,0,0)} \\
\caption{\label{fig:normlayers}Visualizing which axes different normalization layers normalize over. $N$ is the batch dimension, $C$ the channel dimension, and $H,W$ is the spatial dimensions. If the shaded area spans over an axis, this axis is normalized over. For example, BatchNorm normalizes over space and batches, LayerNorm over space and channels. Our normalization layer, ChannelNorm, normalizes over channels only. \emph{Figure adapted from~\cite{wu2018group}.}}
\end{figure}

Fig.~\ref{fig:normlayers} shows a visual comparison of different normalization layers, which is based on the visualization provided in~\cite{wu2018group}.
We note that 
BatchNorm~\cite{ioffe2015batch}, LayerNorm~\cite{ba2016layer}, InstanceNorm~\cite{ulyanov2016instance}, and GroupNorm~\cite{wu2018group}, all average over space.

We compare ChannelNorm to InstanceNorm via the following equations. To simplify notation, we assume that normal broadcasting rules apply, \eg, $f_{chw}-\mu_c$ means that when calculating channel $c$, we subtract $\mu_c$ from each spatial location $h, w$ in the $c$-th channel of $f$. As introduced in Section~\ref{sec:arch}, ChannelNorm uses learned per-channel offsets $\alpha_c, \beta_c$ and normalizes over channels:
\begin{align*}
    f'_{chw} = \frac{f_{chw}-\mu_{hw}}{\sigma_{hw}}
                \alpha_c + \beta_c, 
        &\qquad \mu_{hw} = \sfrac{1}{C} \textstyle \sum_{c=1}^C f_{chw} \\
        &\qquad \sigma_{hw}^2 = \sfrac{1}{C} \textstyle \sum_{c=1}^C (f_{chw} - \mu_{hw})^2.
\end{align*}
InstanceNorm also uses $\alpha_c, \beta_c$, but normalizes spatially:
\begin{align*}
    f'_{chw} = \frac{f_{chw}-\mu_{c}}{\sigma_{c}}
                \alpha_c + \beta_c, 
        &\qquad \mu_{c} = \sfrac{1}{HW} 
          \textstyle \sum_{h=1}^H \textstyle \sum_{w=1}^W 
          f_{chw} \\
        &\qquad \sigma_{c}^2 = \sfrac{1}{HW} 
          \textstyle \sum_{h=1}^H \textstyle \sum_{w=1}^W 
          (f_{chw} - \mu_{c})^2.
\end{align*}
We hypothesize that the dependency on $H, W$ causes the generalization problems we see.

\newpage

\subsection{User Study: More Results} \label{sec:supp:userstudy}

\begin{figure}[ht]
    \centering
    \includegraphics[width=\textwidth]{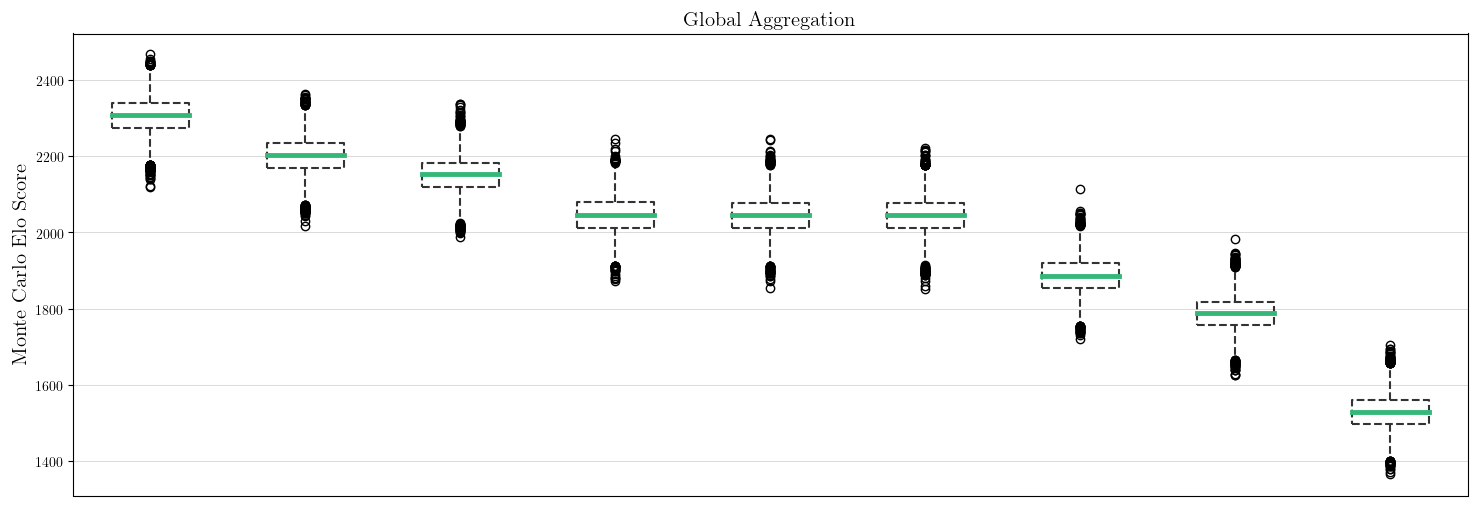}
{\tiny
\begin{tabular}{@{\hskip 0.9cm}p{0.105\textwidth}@{\hskip 0in}p{0.105\textwidth}@{\hskip 0in}p{0.105\textwidth}@{\hskip 0in}p{0.105\textwidth}@{\hskip 0in}p{0.105\textwidth}@{\hskip 0in}p{0.105\textwidth}@{\hskip 0in}p{0.105\textwidth}@{\hskip 0in}p{0.105\textwidth}@{\hskip 0in}p{0.105\textwidth}@{\hskip 0in}p{0.105\textwidth}@{\hskip 0in}}
\namehi \scalebox{.9}[1.0]{\emph{Ours}} & \namemi \scalebox{.9}[1.0]{\emph{Ours}} & BPG & M\&S & \namelo \scalebox{.9}[1.0]{\emph{Ours}} & BPG & M\&S & no GAN & M\&S\\
\textbf{0.359} & \textbf{0.237} & \textbf{0.504} & \textbf{0.405} & \textbf{0.120} & \textbf{0.390} & \textbf{0.272} & \textbf{0.118} & \textbf{0.133}\\
\end{tabular}}
\caption{Global Monte Carlo Elo Scores. We use the standard box plot visualization: The horizontal green thick line indicates the median, the box extends from Q1 to Q3 quartiles, the whiskers extend to $1.5\cdot(Q3-Q1)$, and outlier points are points past the whiskers.}
    \label{fig:userstudy_global}
\end{figure}

For the user study plot in the main text (Fig.~\ref{fig:userstudy}), we averaged Elo scores across all raters, across all images, \ie, the Elo tournament is over all comparisons (see Section~\ref{sec:userstudy}). This data is visualized as a box plot in Fig.~\ref{fig:userstudy_global}.

\begin{figure}[ht]
    \centering
    \includegraphics[width=\textwidth]{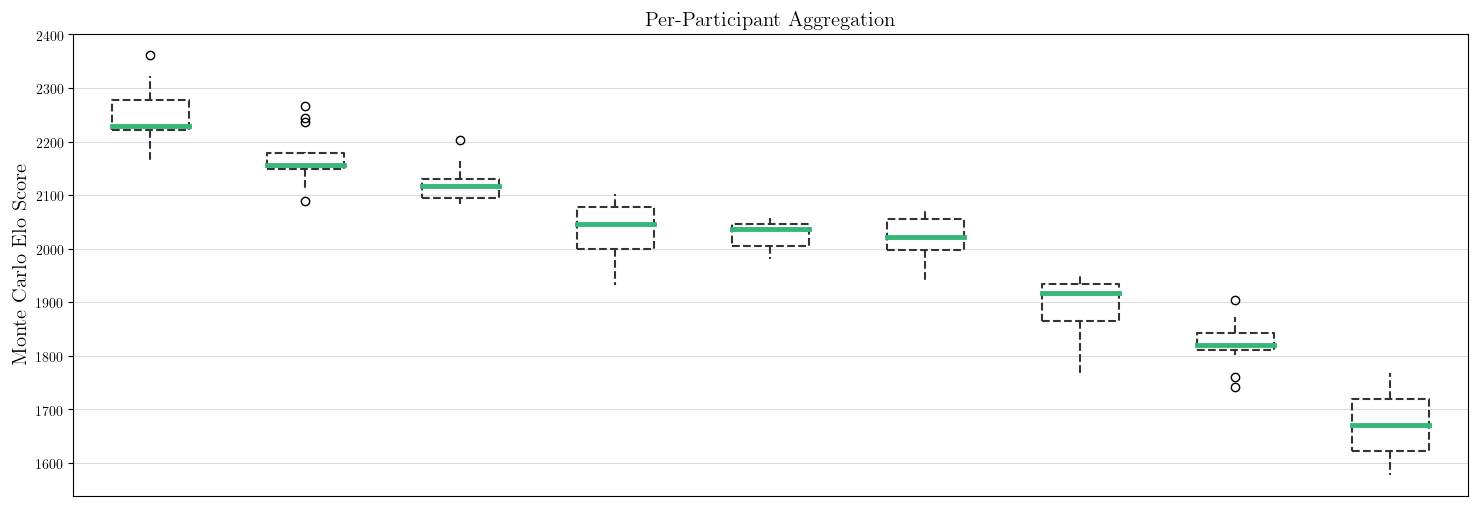}
    {\tiny
\begin{tabular}{@{\hskip 0.9cm}p{0.106\textwidth}@{\hskip 0in}p{0.106\textwidth}@{\hskip 0in}p{0.106\textwidth}@{\hskip 0in}p{0.106\textwidth}@{\hskip 0in}p{0.106\textwidth}@{\hskip 0in}p{0.106\textwidth}@{\hskip 0in}p{0.106\textwidth}@{\hskip 0in}p{0.106\textwidth}@{\hskip 0in}p{0.106\textwidth}@{\hskip 0in}p{0.106\textwidth}@{\hskip 0in}}
\namehi \scalebox{.9}[1.0]{\emph{Ours}} & \namemi \scalebox{.9}[1.0]{\emph{Ours}} & BPG & \namelo \scalebox{.9}[1.0]{\emph{Ours}} & BPG & M\&S & M\&S & no GAN & M\&S\\
\textbf{0.359} & \textbf{0.237} & \textbf{0.504} & \textbf{0.120} & \textbf{0.390} & \textbf{0.405} & \textbf{0.272} & \textbf{0.118} & \textbf{0.133}\\
\end{tabular}}
    \caption{Per-participant Monte Carlo Elo Scores. We use the standard box plot visualization: The horizontal green thick line indicates the median, the box extends from Q1 to Q3 quartiles, the whiskers extend to $1.5\cdot(Q3-Q1)$, and outlier points are points past the whiskers.}
    \label{fig:userstudy_per_rater}
\end{figure}

\begin{figure}[h]
    \centering
    \includegraphics[width=\textwidth]{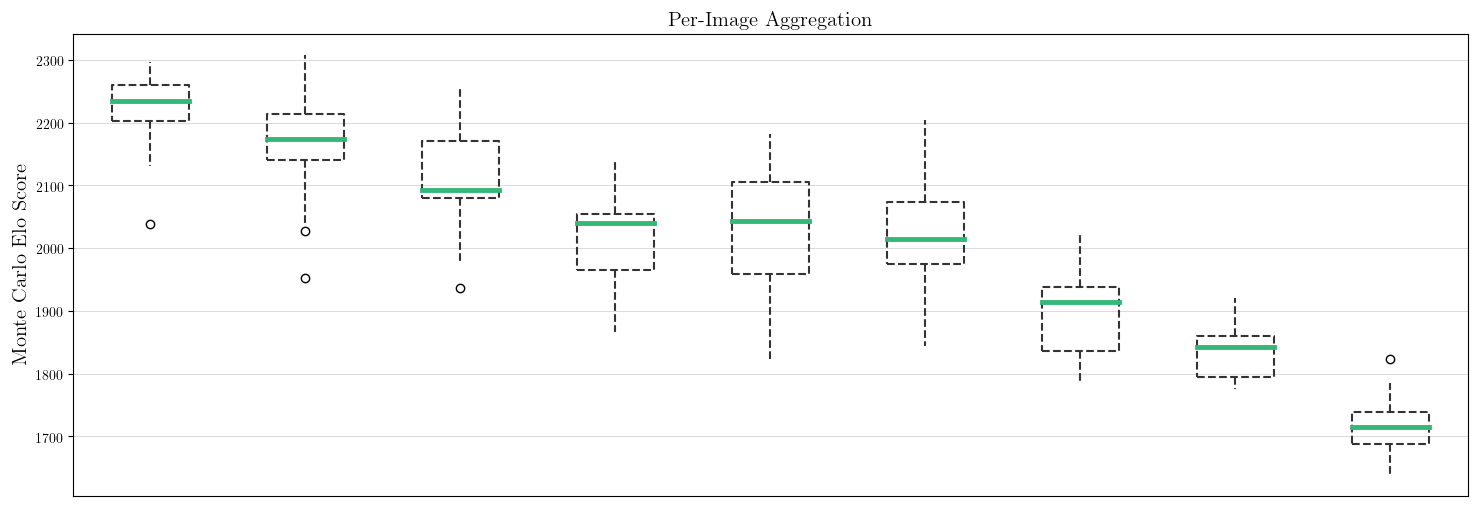}
    {\tiny
\begin{tabular}{@{\hskip 0.9cm}p{0.105\textwidth}@{\hskip 0in}p{0.105\textwidth}@{\hskip 0in}p{0.105\textwidth}@{\hskip 0in}p{0.105\textwidth}@{\hskip 0in}p{0.105\textwidth}@{\hskip 0in}p{0.105\textwidth}@{\hskip 0in}p{0.105\textwidth}@{\hskip 0in}p{0.105\textwidth}@{\hskip 0in}p{0.105\textwidth}@{\hskip 0in}p{0.105\textwidth}@{\hskip 0in}}
\namehi \scalebox{.9}[1.0]{\emph{Ours}} & \namemi \scalebox{.9}[1.0]{\emph{Ours}} & BPG & M\&S & \namelo \scalebox{.9}[1.0]{\emph{Ours}} & BPG & M\&S & no GAN & M\&S\\
\textbf{0.359} & \textbf{0.237} & \textbf{0.504} & \textbf{0.405} & \textbf{0.120} & \textbf{0.390} & \textbf{0.272} & \textbf{0.118} & \textbf{0.133}\\
\end{tabular}}
    \caption{Per-image Monte Carlo Elo Scores. We use the standard box plot visualization: The horizontal green thick line indicates the median, the box extends from Q1 to Q3 quartiles, the whiskers extend to $1.5\cdot(Q3-Q1)$, and outlier points are points past the whiskers.}
    \label{fig:userstudy_per_image}
\end{figure}

In this section, we also show the result of averaging over participants in Fig.~\ref{fig:userstudy_per_rater} (running an Elo tournament per participant), and over images in Fig.~\ref{fig:userstudy_per_image} (running an Elo tournament per image).
As we can see, the overall order remains unchanged, except for the three methods that were very close in median Elo score in Fig.~\ref{fig:userstudy} (\name at 0.120bpp, BPG at 0.390bpp, and \blminnen at 0.405bpp), which change their order. Also, different images obtain a wide range of Elo scores. 

In Table~\ref{tab:per_image}, we show the first 5 characters of the names of the $N_I=20$ images used for the user study, and the rank each method scored on this image, where lower is better. We note that for all images, one of our GAN models earns first or second place -- except for the last image. This image contains a lot of small scale text, and is shown as part of our visualizations in Appendix~\ref{sec:supp:morevisuals}. There, we also provide a link to all images used in the user study.

\begin{figure}[hb]
\vspace{-4em}
    \centering
    \includegraphics[width=0.8\textwidth]{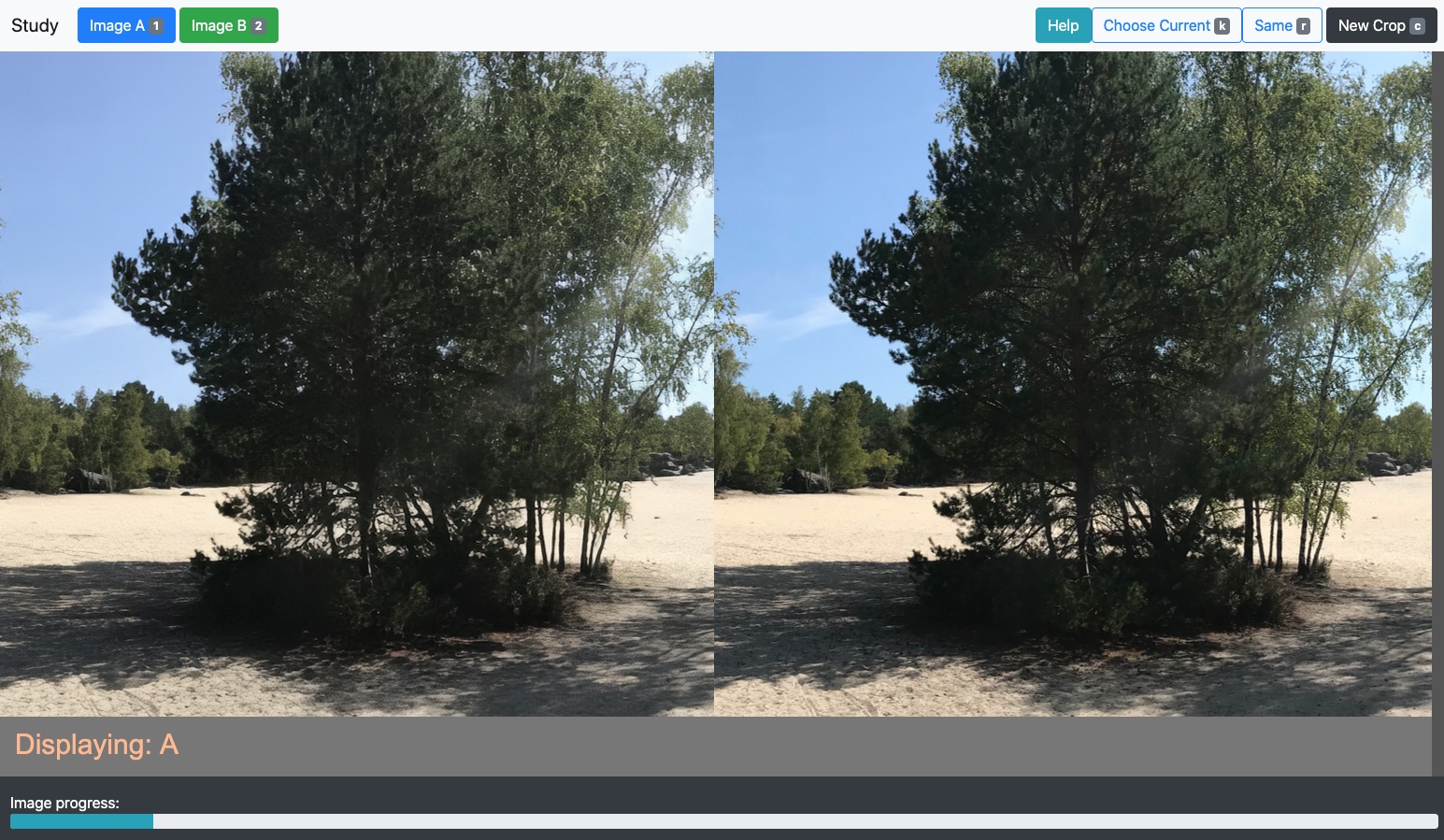}
    \caption{\label{fig:userstudy_gui}Screenshot of the user study GUI.}
    \vspace{0.5ex}  
\end{figure}

In Fig.~\ref{fig:userstudy_gui}, we show a screenshot of the GUI shown to raters. The crops selected by the participants are available at \url{https://hific.github.io/raw/userstudy.json}.

\begin{table}[h]
\scriptsize
\center
\begin{tabular}{lccccccccc}
\toprule
\rotatebox{90}{Image name} 
& \rotatebox{90}{\namehi\, 0.359 bpp} & \rotatebox{90}{\namemi\, 0.237 bpp} & \rotatebox{90}{BPG\, 0.504 bpp} & \rotatebox{90}{M\&S\, 0.405 bpp} & \rotatebox{90}{\namelo\, 0.120 bpp} & \rotatebox{90}{BPG\, 0.390 bpp} & \rotatebox{90}{M\&S\, 0.272 bpp} & \rotatebox{90}{no GAN\, 0.118 bpp} & \rotatebox{90}{M\&S\, 0.133 bpp} \\
\midrule
\texttt{e0256}  & 1 & 2 & 3 & 5 & 4 & 6 & 8 & 7 & 9 \\
\texttt{a251f}  & 1 & 2 & 4 & 8 & 3 & 5 & 7 & 6 & 9 \\
\texttt{0ae78}  & 1 & 2 & 3 & 6 & 5 & 4 & 8 & 7 & 9 \\
\texttt{95e7d}  & 1 & 2 & 3 & 6 & 4 & 5 & 8 & 7 & 9 \\  
\texttt{2145f}  & 1 & 2 & 4 & 3 & 5 & 6 & 7 & 8 & 9 \\
\texttt{58c13}  & 2 & 1 & 3 & 5 & 6 & 4 & 7 & 9 & 8 \\
\texttt{f063e}  & 2 & 4 & 1 & 6 & 5 & 3 & 7 & 8 & 9 \\
\texttt{dcb53}  & 1 & 2 & 4 & 3 & 5 & 6 & 7 & 8 & 9 \\
\texttt{d5424}  & 1 & 2 & 3 & 5 & 6 & 4 & 8 & 7 & 9 \\
\texttt{72e19}  & 1 & 2 & 5 & 4 & 3 & 8 & 6 & 7 & 9 \\
\texttt{1c55a}  & 1 & 5 & 2 & 4 & 6 & 3 & 7 & 8 & 9 \\
\texttt{ad249}  & 2 & 1 & 5 & 4 & 3 & 7 & 8 & 6 & 9 \\  
\texttt{d9692}  & 2 & 5 & 1 & 4 & 6 & 3 & 7 & 8 & 9 \\
\texttt{18089}  & 1 & 3 & 2 & 5 & 6 & 4 & 7 & 8 & 9 \\
\texttt{f7a9e}  & 1 & 2 & 4 & 3 & 5 & 6 & 7 & 8 & 9 \\
\texttt{a09ce}  & 1 & 3 & 2 & 5 & 7 & 4 & 6 & 8 & 9 \\
\texttt{6e8e3}  & 1 & 2 & 4 & 5 & 3 & 6 & 8 & 7 & 9 \\
\texttt{afa0a}  & 1 & 2 & 4 & 6 & 3 & 5 & 7 & 8 & 9 \\
\texttt{8ba19}  & 2 & 1 & 3 & 5 & 4 & 6 & 7 & 8 & 9 \\
\texttt{25bf4}  & 4 & 6 & 1 & 3 & 8 & 2 & 5 & 7 & 9 \\
\bottomrule
\end{tabular}
\caption{\label{tab:per_image} Per image rankings of the user study. We show the average bpp of each method in the header}
\end{table}
\FloatBarrier

\subsection{Training Details} \label{sec:supp:training}  
We follow standard practice of alternating between training $E$, $G$, $P$ (jointly) for one step and training $D$ for one step. We use the same learning rate of $1\textsc{e}^{-4}$ for all networks, and train using the Adam optimizer~\cite{kingmaB14}.
At the beginning of training, the rate loss can dominate. To alleviate this, we always first train with a higher $\lR$, as in~\cite{minnen2018joint}.
As mentioned in Section~\ref{sec:evaluation}, we initialize our GAN models (\namehi, \namemi, \namelo) from a model trained for MSE and $d_P=$LPIPS. 
Table~\ref{tab:schedules} shows our different models and the LR and $\lR$ schedules we use. The GAN initialization is using the \emph{Warmup} model. Together, this yields 2M steps for all models.
\begin{table}[h]
\begin{center}
\small
\begin{tabular}{llcccc}
\toprule
 & Losses & Initialize with & Training & LR decay & Higher $\lR$ \\
\midrule
\blmselpips   & MSE+LPIPS & - & 2M steps & 1.6M steps & 1M steps \\
\blminnen     & MSE       & - & 2M steps & 1.6M steps & 1M steps \\[3pt]
\emph{Warmup} & MSE+LPIPS & - &1M steps & 0.5M steps & 50k steps \\
\name\textsuperscript{Hi,Mi,Lo}         & MSE+LPIPS+GAN & \emph{Warmup} & 1M steps & 0.5M steps & 50k steps \\
\bottomrule
\end{tabular}
\vspace{1ex}
\caption{\label{tab:schedules}Training schedules, using ``M'' for million, ``k'' for thousand.}
\end{center}
\end{table}

We fix hyper-parameters shown in Fig.~\ref{fig:hyper:fixed} for all experiments (unless noted), and vary $\lRa$ depending on $\rtarget$ as shown in Fig.~\ref{fig:hyper:var}.

The training code and configs for \enamelo and \eblmselpips is available at 
\href{https://hific.github.io}{\textbf{\texttt{hific.github.io}}}.

\begin{figure}[h]
\vspace{-1em}
\centering
\small
\begin{subfigure}[b]{0.49\textwidth}
\centering
\hfill\begin{tabular}{lll}
\toprule
    Parameter & Value & Note \\
\midrule
    $\lRb$   & $2^{-4}$ \\
    $\lMSE$  & $0.075 \cdot 2^{-5}$ \\
    $\lP$    & 1 \\
    $C_y$    & 220 \\
    $\beta$  & $0.15$ & Except for Fig.~\ref{fig:cp} \\
\bottomrule
\end{tabular}
\caption{\label{fig:hyper:fixed} Fixed hyper-parameters.}
\end{subfigure}
\begin{subfigure}[b]{0.3\textwidth}
\centering
\begin{tabular}{ll}
\toprule
     Target Rate $\rtarget$ & $\lRa$  \\
\midrule
     0.14 & $2^1$ \\
     0.30 & $2^0$ \\
     0.45 & $2^{-1}$ \\
\bottomrule
\end{tabular}\hfill%
\caption{\label{fig:hyper:var}Varying hyper-parameters.}
\end{subfigure}
\vspace{-0.5ex}
\caption{Hyper-parameters.}
\end{figure}

\subsection{Patch-based FID and KID} \label{sec:supp:fidpatches}
\vsqueeze\vsqueezehalf
As mentioned in Section~\ref{sec:evaluation}, we extract patches to calculate FID and KID. From each $H{\times}W$ image, we first extract $\lfloor H/f \rfloor \cdot \lfloor W/f \rfloor$ non-overlapping $f\times f$ crops, and then shift the extraction origin by $f{/}2$ in both dimensions to extract another 
$(\lfloor H/f \rfloor - 1)\cdot (\lfloor W/f \rfloor - 1)$ patches. We use $f=256$ in all evaluations.

\subsection{Further Quantitative Results} \label{sec:supp:furthereval}
\vsqueeze\vsqueezehalf

\begin{figure}
    \centering
    \includegraphics[width=\textwidth]{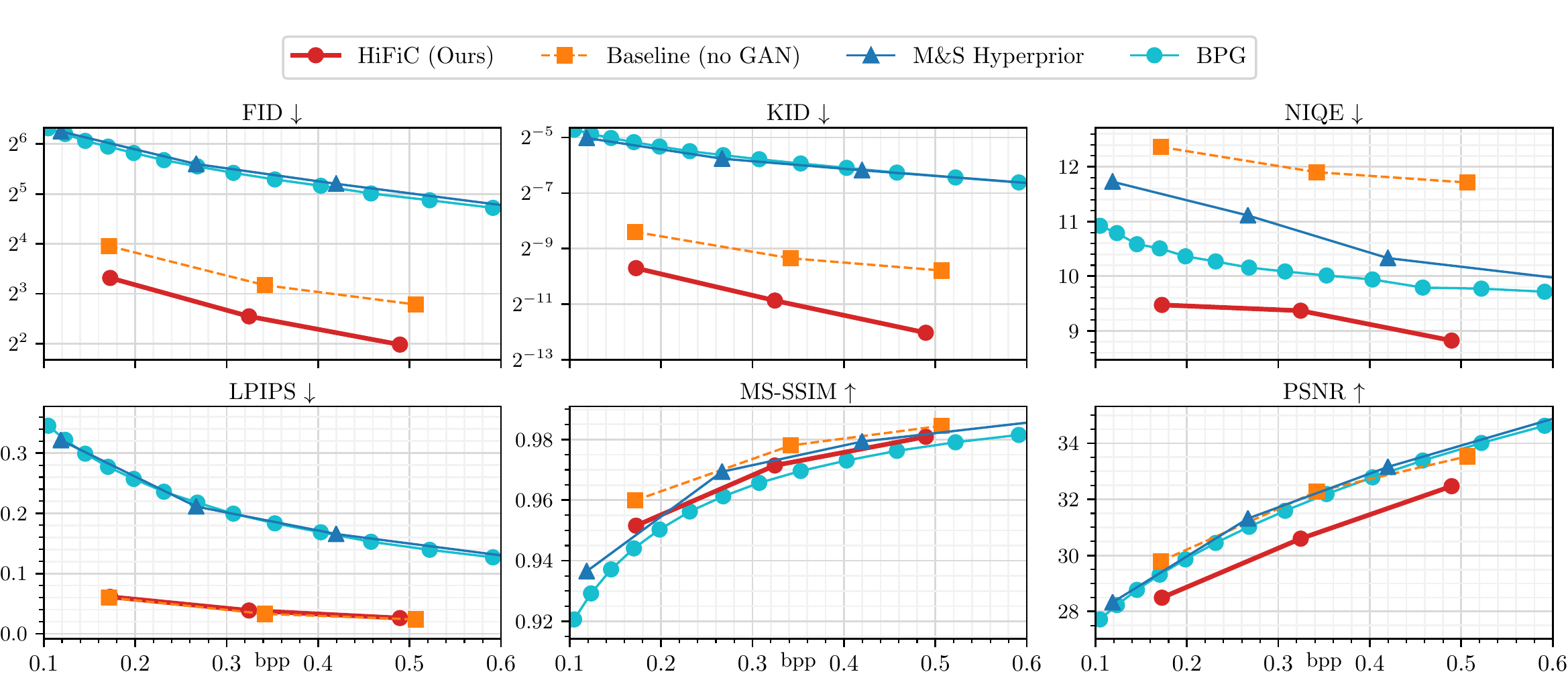}
    \vspace{-3ex}
    \caption{Rate-distortion and -perception curves on DIV2K. Arrows in the title indicate whether lower is better ($\downarrow$), or higher is better ($\uparrow$).} 
    \label{fig:rd_rp_div2k}
\end{figure}

\begin{figure}
\vspace{-2em}
    \centering
    \includegraphics[width=0.55\textwidth]{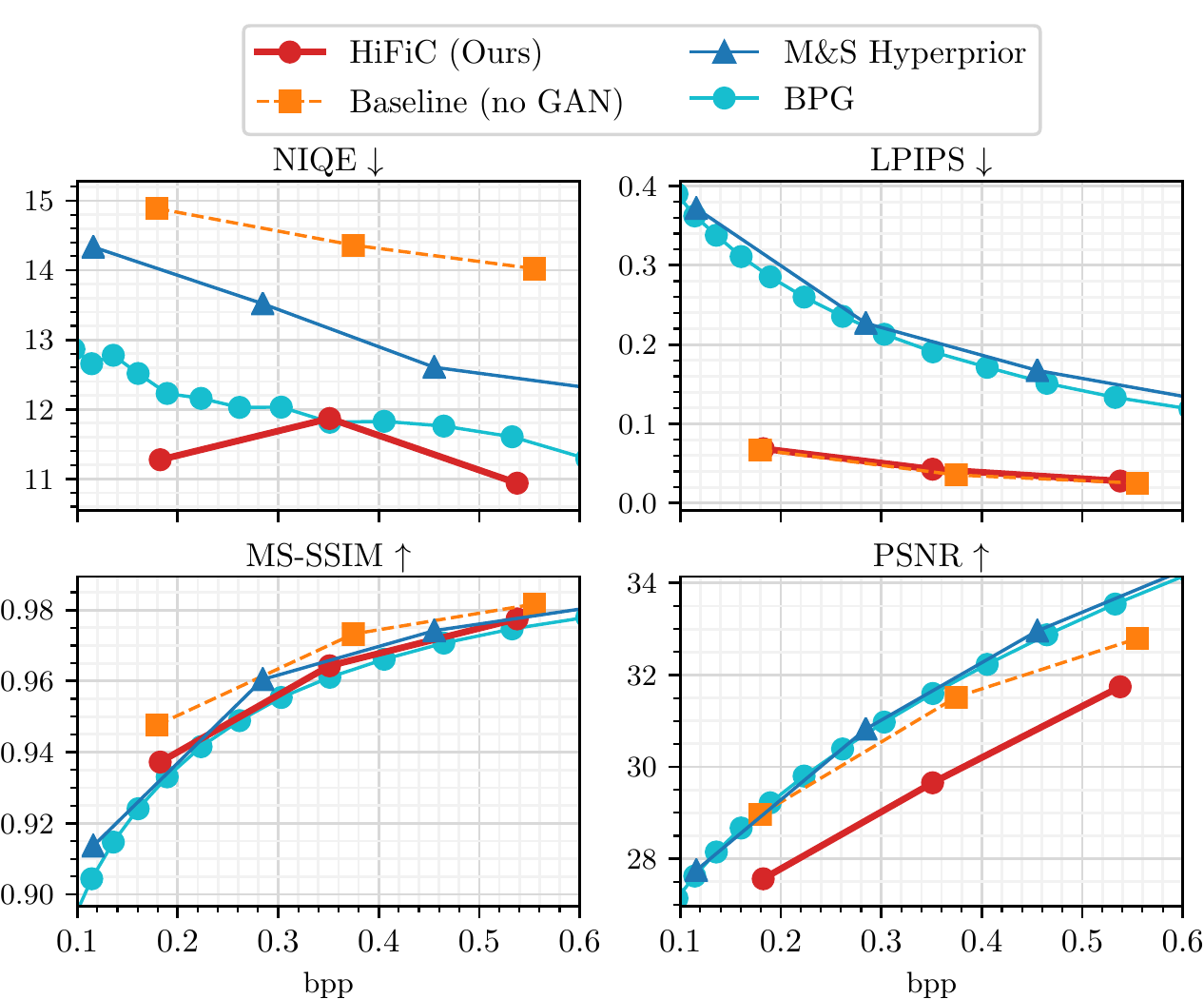}
    \vspace{-1ex}
    \caption{Rate-distortion and -perception curves on Kodak. Arrows in the title indicate whether lower is better ($\downarrow$), or higher is better ($\uparrow$).} 
    \label{fig:rd_rp_kodak}
\end{figure}

In this section, we provide plots similar to Fig.~\ref{fig:rd_rp} on the other two datasets:
In Fig.~\ref{fig:rd_rp_div2k}, we show rate-distortion and rate-perception curves for DIV2K~\cite{somanywaystocitediv2k}, and in Fig.~\ref{fig:rd_rp_kodak}, we show curves for Kodak~\cite{kodakurl}.
As noted in Section~\ref{sec:evaluation}, the 24 Kodak images only yield 192 patches to calculate FID and KID, and we thus omit the two metrics.

\subsection{Image Dimensions of the Datasets}  \label{sec:supp:dataset}
\begin{figure}
    \centering
    \includegraphics[width=0.8\textwidth]{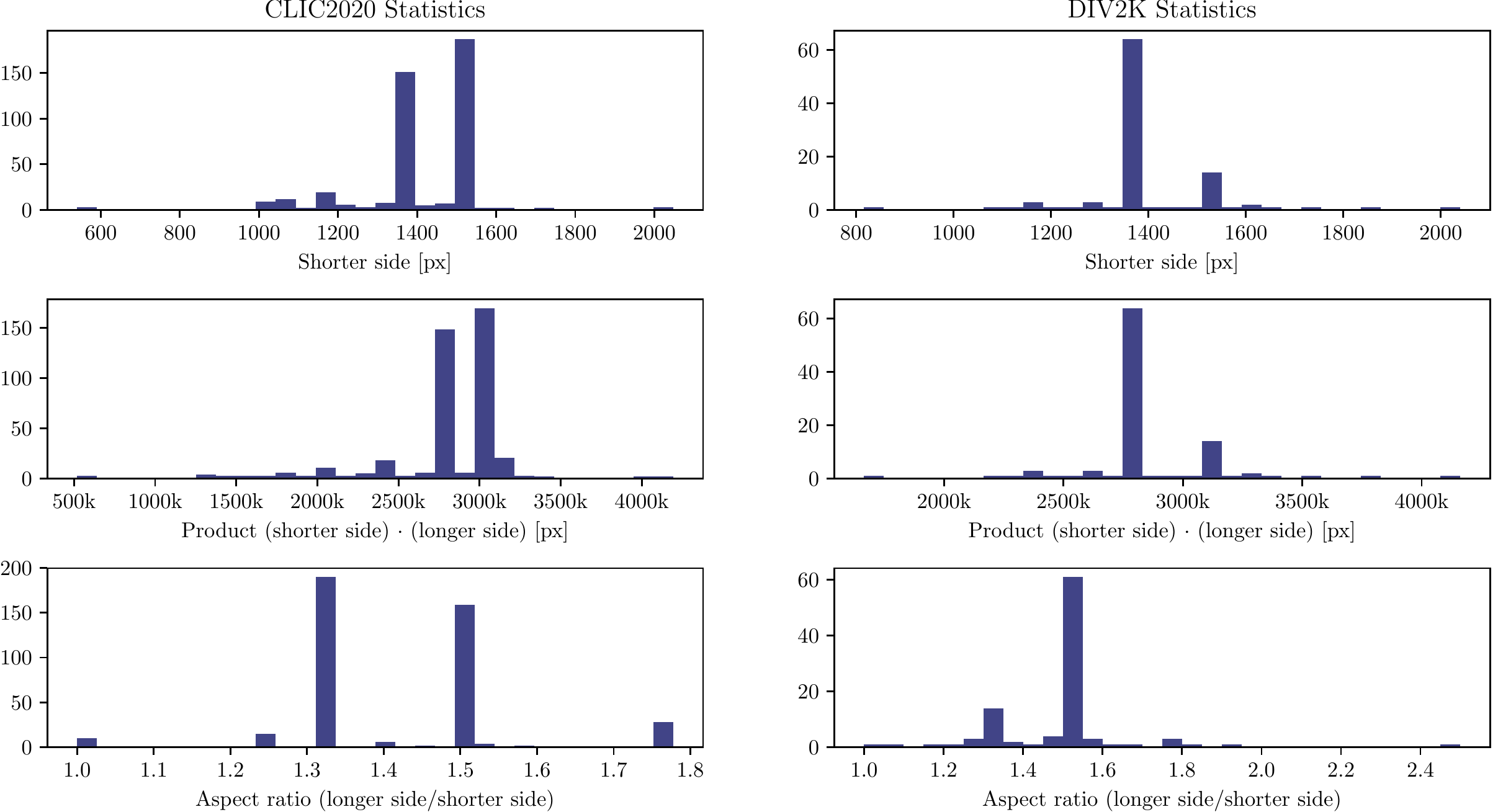}
    \caption{\label{fig:dataset_stat}Histograms to show image dimensions for CLIC2020 and DIV2K.}
\end{figure}

\vsqueeze\vsqueezehalf
We use the three datasets mentioned in Section~\ref{sec:evaluation} for our evaluation. Kodak contains images of $768{\times}512$ pixels, the other two datasets contain images of varying dimensions. To visualize the distribution of these dimensions, we show histograms for the shorter dimensions, for the product of both dimensions, and for the aspect ratio in Fig.~\ref{fig:dataset_stat}.
We see that most images cluster around shorter sides of 1400px, and go up to 2000px. We note that the three biggest images for CLIC are of dimensions $2048{\times}2048, 2048{\times}2048, 2000{\times}2000$, for DIV2K they are $2040{\times}2040, 1872{\times}2040, 1740{\times}2040$.

\FloatBarrier
\clearpage

%
%

\section{Further Visual Results} \label{sec:supp:morevisuals}

\subsection{PDF Visualization}

\emph{Due to size constraints, the PDF is hosted at} \url{https://hific.github.io/appendixb}.
                      
In the PDF, we show images from all datasets. For each image, we show the full reconstruction by one of our models, next to the original. On the top left of this image, we show the dataset and image ID.
Additionally, we pick one or two crops for each image, where we compare the original to the following methods:
\begin{enumerate}[topsep=0pt]
    \item \nameoursmi
    \item \nameourslo
    \item \eblminnen at a bitrate greater than \enamelo
    \item BPG at the $2{\times}$ the bitrate of \enamemi
    \item BPG at the same bitrate as \enamemi
    \item BPG at the same bitrate as \enamelo
    \item JPEG at $Q=80$.
\end{enumerate}
We chose to add JPEG at $Q=80$ as a further reference, as this is a quality factor in common use~\cite{coulombe2009low}. 

\subsubsection*{Results}

Throughout the examples, we can see that our GAN models shine at reconstructing plausible textures, yielding reconstructions that are very close to the original. We see that BPG at the same bitrate as \enameourslo tends to exibit block artifacts, while \enamelo looks very realistic. For most images, our \enamemi model also looks significantly better than BPG at the same bitrate. When BPG uses $2{\times}$ the rate as \enamemi, it starts to look similar to our reconstructions.

\subsection{More Comparisons and Raw Images}

We note that the crops are embedded as PNGs in the PDF, but the large background images are embedded as JPEGs to prevent a huge file. However, we package full-sized PNGs for various methods into ZIPs. The ZIP files also contain a HTML file that can be used for easy side-by-side comparisons, as this is the best way to visualize differences. 

The images are also available directly in the browser at:

\url{https://hific.github.io/raw/index.html}.

The ZIP files:
\begin{enumerate}[topsep=0pt,itemsep=3mm]
    \item The following ZIP contains the 20 images from CLIC2020 used for the user study, compressed with each of the 9 methods used in the study. \vspace{1mm}\\
    \url{https://hific.github.io/raw/userstudy} \\
    Size: 610MB 
    \item The following ZIP files contain all the images of the respective datasets, compressed with \enamehi, \enamemi, \enamelo. \vspace{1mm} \\
    \url{https://hific.github.io/raw/kodak} \\
    Size: 61MB \\
    \url{https://hific.github.io/raw/clic2020} \\
    Size: 6.3GB \\
    \url{https://hific.github.io/raw/div2k} \\
    Size: 1.7GB
\end{enumerate}

\end{document}